\documentclass[twocolumn,showpacs,preprintnumbers,amsmath,amssymb]{revtex4}

\usepackage{graphicx}
\usepackage{dcolumn}
\usepackage{bm}
\usepackage{float}
\usepackage[normalem]{ulem}
\usepackage{color}
\usepackage{refstyle}
\usepackage{mathrsfs}
\usepackage{esint}
\usepackage[colorinlistoftodos]{todonotes}
\usepackage[unicode=true,pdfusetitle,bookmarks=true,
bookmarksnumbered=false,bookmarksopen=false,breaklinks=false,
pdfborder={0 0 0},backref=false,colorlinks=true,citecolor=red]
{hyperref}
\usepackage{cleveref}
\providecommand\eqref[1]{\ref{eq:#1}}
\renewcommand\b[1]{{\bf  #1}}
\renewcommand\epsilon{\varepsilon}
\renewcommand\vec[1]{\boldsymbol{#1}}
\renewcommand\phi{\varphi}

\newcommand\del{\nabla}
\newcommand\dd{\mathrm{d}}
\allowdisplaybreaks[3]

\begin{document}
\title{The low noise phase of a $2d$ active nematic}
\author{Suraj Shankar$^{a,b}$}
\author{Sriram Ramaswamy$^{c}$}
\author{M. Cristina Marchetti$^{a}$}
\affiliation{$^a$Physics Department and Syracuse Soft Matter Program, Syracuse University, Syracuse, NY 13244, USA.\\
$^b$Kavli Institute for Theoretical Physics, University of California, Santa Barbara, CA 93106, USA.\\
$^c$Centre for Condensed Matter Theory, Department of Physics, Indian Institute of Science, Bangalore 560012, India.}

\date{\today}
\begin{abstract}
	We consider a collection of self-driven apolar particles on a substrate that organize into an active nematic phase at sufficiently high density or low noise. Using the dynamical renormalization group, we systematically study the $2d$ fluctuating ordered phase in a coarse-grained hydrodynamic description involving both the nematic director and the conserved density field. In the presence of noise, we show that the system always displays only quasi-long ranged orientational order beyond a crossover scale. A careful analysis of the nonlinearities permitted by symmetry reveals that activity is \emph{dangerously irrelevant} over the linearized description, allowing giant number fluctuations to persist though now with strong finite-size effects and a non-universal scaling exponent. Nonlinear effects from the active currents lead to power law correlations in the density field thereby preventing macroscopic phase separation in the thermodynamic limit.
\end{abstract}
\maketitle
\section{Introduction} \label{intro}
Collections of self-propelled units that are driven out of equilibrium by the consumption of free energy at the microscopic level spontaneously organize in a variety of active matter states \cite{marchetti2013hydrodynamics,ramaswamy2010mechanics}. When elongated in shape, such units form active liquid crystalline phases that may have polar or nematic symmetry. An active nematic is by far the simplest realization of an active system that can display orientational order. Unlike its polar counterpart, where the appearance of macroscopic polar order results in collective directed motion or flocking \cite{toner2005hydrodynamics,toner1998flocks}, the active nematic involves driven \emph{apolar} constituents, which means on average the system goes nowhere \cite{ramaswamy2003active} making its properties far more subtle. Examples of active nematics include monolayers of melanocytes \cite{kemkemer2000elastic,gruler1999nematic}, fibroblasts \cite{duclos2014perfect}, neural progenitors \cite{kawaguchi2017topological}, myxobacteria \cite{starruss2012pattern,thutupalli2015directional}, swimming filamentous bacteria \cite{PhysRevE.95.020601,zhou2014living,genkin2017topological}, vibrated rods \cite{narayan2007long} and microtubule-kinesin suspensions \cite{sanchez2012spontaneous}.

The theoretical study of active nematics began with coarse-grained approaches \cite{ramaswamy2003active,simha2002hydrodynamic}, followed by numerical agent-based \cite{chate2006simple,ngo2014large} or lattice gas simulations \cite{mishra2006active} of minimal microscopic models. In two dimensions ($2d$), numerical work by \citet{ngo2014large} revealed an order-disorder transition that involved three phases - (i) a homogeneous disordered gas at high noise and low density, (ii) an intermediate locally banded, chaotic, macroscopically isotropic but segregated phase, and (iii) a homogeneous but fluctuating (quasi)-ordered nematic phase at low noise and high density. The segregated phase is presumably a result of the instability of the homogeneous nematic phase to band formation close to the mean-field transition \cite{shi2010deterministic,putzig2014phase,shi2014instabilities,bertin2013mesoscopic}. The lines delimiting the chaotic band phase determine the binodal lines. The linear instability of the ordered phase then corresponds to the spinodal which falls well within the band forming region. The inhomogeneous bands are themselves unstable to transverse fluctuations (in a large enough system), leading to the intermediate chaotic and phase-separated but isotropic phase between the binodals. This should be contrasted with the polar case, where a spatially periodic phase of coherently moving stable bands is seen just past the flocking transition \cite{solon2015phase}. An analytical understanding of the transition from the chaotic biphasic state to the ordered nematic phase is unavailable, and  strong density fluctuations obscure its character even in numerical studies \cite{ngo2014large}. In ``metric-free'' models, in which the interaction neighbourhood is the first Voronoi shell, numerical studies \cite{ngo2014large} find only two phases, both homogeneous : a quasi-long-range ordered nematic and an isotropic phase, separated by a transition of Berezinskii-Kosterlitz-Thouless type \cite{berezinskii1971destruction,kosterlitz1973ordering}. There has also been a lot of previous work at the continuum level (in the absence of noise) on ``wet'' active nematic systems, i.e., including flow and hydrodynamic interactions~\cite{hemingway2016correlation,thampi2014vorticity,thampi2013velocity,thampi2014instabilities,giomi2013defect,giomi2011excitable,giomi2015geometry}.

Giant number fluctuations (GNFs) \cite{toner1998flocks,ramaswamy2003active,toner2005hydrodynamics,narayan2007long} are a ubiquitous property of the orientationally ordered phases of active systems. As emphasized in Ref.~\cite{PhysRevE.95.020601}, it is important to distinguish GNFs from regular phase separation, generically present close to the transition, as well as from the inhomogeneous structures that occur in fluctuation-dominated transitions \cite{das2000particles}. A study of the ideal phenomenology of these anomalous fluctuations requires a well developed ordered phase in a large enough system which has a mean \emph{homogeneous} density and is \emph{not} phase separated in the thermodynamic limit.
%
Here we examine the stability of \emph{any} ordered active nematic phase to the introduction of noise. A previous dynamical renormalization group analysis of a $2d$ active nematic on a substrate in the \emph{absence} of a conserved density \cite{mishra2010dynamic} showed that anisotropic nonlinearities, including a contribution from advection by active currents, are perturbatively irrelevant in the infrared, leading to an equilibrium XY model like description at long-distances, and hence quasi-long-ranged order (QLRO) at low noise. Here we take on the more ambitious program of reinstating the density field in the RG analysis to establish the behavior of both orientational and density fluctuations in the nematic phase.

The main results of our work are summarized as follows:
\begin{itemize}
	\item\emph{Quasi-long range order (QLRO) in $2d$ active nematics:} We show that in a system of linear size $L$ and small-scale cutoff $a$ the nematic order parameter $\langle\Psi_{2}\rangle$ asymptotically decays as
\begin{equation}
\langle\Psi_{2}\rangle\sim \left(\frac{L}{a}\right)^{-\eta(\Delta)}\ ,
\label{eq:r1}
\end{equation}
where $\eta(\Delta)=\Delta/2\pi K$ is a nonuniversal exponent varying continuously with the noise strength $\Delta$ and the nematic stiffness $K$. Thus the quasi-long-range order found for active nematics in the absence of number conservation \cite{mishra2010dynamic} continues to hold upon introduction of a locally conserved number-density field. Note that in equilibrium both polar and nematic liquid crystals behave at long wavelengths like an XY model in $2d$. \emph{Active} polar and nematic systems are, however, distinct. A $2d$ active polar fluid exhibits LRO \cite{toner1995long}, while $2d$ active nematics, like equilibrium ones, exhibit only QLRO. Moreover, the exponent $\eta(\Delta)$ has the same form as that of an equilibrium XY model with the noise $\Delta$ taking the role of temperature.
\item\emph{Giant Number Fluctuations (GNFs):} The power-law decay of the order parameter yields an associated power-law scaling of density fluctuations. As a result, the standard deviation $\Delta N$ of particle number in a region containing on average $N$ particles is found to scale as 
\begin{equation}
\Delta N \sim N^{1 - \eta(\Delta)/2}\;.
\label{eq:r2}
\end{equation}
Note that a mean-field analysis yields $\eta(\Delta)=0$ in $2d$ \cite{ramaswamy2003active}. This nonuniversal scaling is a result of marginally, but dangerously \cite{amit1982dangerous}, irrelevant nonlinearities in the active current, and offers a possible explanation for the density fluctuation spectrum observed in the numerical studies of Ngo \textit{et al.} who obtain $\eta(\Delta)=0.4$ \cite{ngo2014large}. In our theory, however, the weakened GNFs (\ref{eq:r2}) are determined by the same exponent as that governing quasi-long-range order (\ref{eq:r1}). \citet{ngo2014large} report a greater suppression of GNFs than can be accounted for by their small observed values of the QLRO exponent. We have no explanation at present for this disagreement between theory and observation.
\item\emph{Strong finite-size effects at large activity:} At high effective activity QLRO as given in Eq.~\ref{eq:r1} is seen only for $L > \xi_* \simeq a \exp[\bar{\lambda}_0^{4/9}(\pi K_0 / \Delta)^{13/9}]$ where $K_0$ is the \textit{bare} nematic stiffness and $\bar{\lambda}_0$ is the \textit{bare} value of a non-dimensional active drive. There is a broad range of system sizes, $a \ll L \ll \xi_*$, where the effective stiffness grows as $[\ln(L/a)]^{4/13}$, and the nematic order parameter thus decreases more slowly than any power of $L$. Simulations or experimental realizations probing a limited range of scales could thus give the impression of long-range order.
\item\emph{GNFs versus phase separation:} We show explicitly that GNFs are distinct from phase separation, even when the latter is induced or dominated by fluctuations~\cite{mishra2006active,das2000particles}. At large activities and on scales smaller than $\xi_*$, we find, however, $\Delta N \sim N (\ln N)^{-5/13}$, which could mimic phase separation to some degree.
\end{itemize}

The remainder of the paper is organized as follows. In Sec.~\ref{sec:model}, we describe the continuum model for a general $2d$ active nematic on a substrate. The reduction of the dynamics to just the slow fields, relevant to the ordered phase is done in Sec.~\ref{subsec:thetarho}. In Sec.~\ref{sec:linhydro} we briefly discuss the linearized hydrodynamic theory and assess the importance of nonlinearities. In analogy with $3d$ fully developed Navier-Stokes turbulence \cite{eyink1994renormalization}, we find that the ordered phase of an active nematic is controlled by an infinite spectrum of marginal operators perturbing the linearized description. At the nonlinear level, in Sec.~\ref{subsec:symmetry}, we analyze the constraints imposed by rotational symmetry and parity at the level of the dynamical equations. It is here that the nematic or apolar nature of the order plays an important role, distinguishing itself from its polar counterpart. In Sec.~\ref{sec:drg}, we perform a low noise expansion about the homogeneous and uniformly ordered state within the framework of the dynamical renormalization group. We emphasize the crucial role of symmetries in allowing us to systematically analyze the infinite tower of nonlinear terms and show to leading order that nearly all of them are marginally irrelevant. We also analyze the flow diagram and show that at long wavelengths only quasi-long ranged nematic order survives in the system, though with possibly very strong finite size and crossover effects. Finally in Sec.~\ref{sec:phasesep}, we address the nature of phase separation in light of the modified giant number fluctuation scaling.

\section{The Model}
\label{sec:model}
We consider a $2d$ active nematic fluid on a frictional substrate. Working at the continuum level, we only have two relevant fields, one is the density ($\rho$) and the other is the nematic alignment tensor $Q_{ij}=S(\hat{n}_i\hat{n}_j-\delta_{ij}/2)$. We rule out topological defects by fiat and conduct only a ``spin-wave'' analysis. This is not merely to avoid technical difficulties but also because the numerical studies of \citet{ngo2014large} find an active nematic phase free of defect proliferation. The scalar order parameter $S$ vanishes in the disordered phase, while $S\neq0$ in the ordered nematic with the direction of broken symmetry given by the director $\hat{\b{n}}$. Particle number conservation implies that the density obeys a continuity equation
\begin{equation}
	\partial_t\rho+\del\cdot\b{j}=0\ ,
\end{equation}
with mass current $\b{j} = \rho \b{u}$ where $\b{u}$ is the velocity field. As the substrate is a momentum sink, $\b{j}$ is itself a fast mode, slaved to variations in $\rho$ and $\b{Q}$. In general the fluctuating current $\b{j}=-M\del\mu+\b{j}_{\mathrm{curl}}+\b{f}_{\rho}$, where for simplicity we have taken a scalar mobility ${M}$, the scalar $\mu$ is an effective chemical potential, $\b{j}_{\mathrm{curl}}$ includes all non-potential contributions to the mass current ($\del\times\b{j}_{\mathrm{curl}}\neq0$), and $\b{f}_{\rho}$ is a gaussian white noise accounting for fluctuations. For a passive system that relaxes to thermal equilibrium with probability distribution $\exp(-\mathcal{F}/k_B T)$, $\mu=\delta\mathcal{F}/\delta\rho$,
$\b{j}_{\mathrm{curl}}=\nu \del\cdot\b{\sigma}^e$ ($\b{\sigma}^e \simeq\delta\mathcal{F}/\delta\b{Q}$ is the field thermodynamically conjugate to the liquid crystal order parameter and $\nu$ is a dissipative cross-coupling) and correlations of $\b{f}_{\rho}$ are related to ${M}$ and the temperature by the fluctuation-dissipation theorem \footnote{Including a dependence of fields in the mobility ${M}$ makes the noise multiplicative, which can sometimes require additional currents $\sim\mathcal{O}(k_BT)$ that must be included to ensure detailed balance \cite{kim1991equations,lau2007state,van1992stochastic}. We shall neglect such complications at present.}. Active contributions breaking detailed balance arise in all three terms, with a non-integrable addition to $\mu$ as in \emph{scalar} active matter \cite{wittkowski2014scalar} and to $\b{j}_{\mathrm{curl}}$ (relevant to active aligning matter), and a violation of the fluctuation-dissipation relation. The leading terms in a gradient expansion are
\begin{gather}
	\mu=c_0\delta\rho+c_1\delta\rho^2+c_2\del^2\delta\rho+2c'_2\del_i\del_jQ_{ij}\ ,\label{eq:chempot}\\
	{M}=(M_1+M'_1\delta\rho)\ ,\\
	\b{j}_{\mathrm{curl}}=\alpha_0\del\cdot\b{Q}+\alpha_1\delta\rho\del\cdot\b{Q}\ ,
\end{gather}
where $\delta\rho=\rho-\rho_0$ is the deviation of the density from its mean and $\b{I}$ is the unit tensor. The most relevant active contribution is the curvature induced current $\propto\alpha_0\del\cdot\b{Q}$ \cite{ramaswamy2003active}, that permits circulating probability currents even in the steady state. 
All the terms included in $\mu$ are present in equilibrium too. The simplest active contribution to the chemical potential $\sim|\del\delta\rho|^2$ is irrelevant (along with other equilibrium terms like $c_2$ and $c'_2$ given above) at long wavelengths. We also only consider systems that are stable and non-phase separating in the absence of activity, so $c_0$ and $M_{1}>0$.

The nematic alignment tensor is not a conserved field and its dynamics is of relaxational form,
\begin{align}
	\partial_t&Q_{ij}=\left(a-\dfrac{b}{2}|\b{Q}|^2\right)Q_{ij}+K\del^2Q_{ij}+L_1Q_{k\ell}\del_k\del_{\ell}Q_{ij}\nonumber\\
												  &+L_2\del_kQ_{k\ell}\del_{\ell}Q_{ij}-2\lambda\left(\del_i\del_j-\dfrac{\delta_{ij}}{2}\del^2\right)\delta\rho+f_{\b{Q}ij}\ ,\label{eq:Qeq}
\end{align}
with the rotational viscosity set to unity by rescaling the unit of time.
The terms $a,b$ account for symmetry breaking allowing for the mean field isotropic-nematic transition. To linear order $a=a_0+a_1\delta\rho$ and $b=b_0+b_1\delta\rho$ with $a_{0}>0$ and $b_{0}>0$ well in the ordered nematic phase. The form of the equation is the same for an equilibrium passive nematic liquid crystal, except that at equilibrium $L_1$, $L_2$ and $K$ (along with terms schematically of the form $\sim\del\b{Q}\del\b{Q}$ with all possible index contractions) would have been related via the free energy to the two independent Frank elastic constants in $2d$.

We digress briefly to dispose of a possible confusion. For a passive system, the gaussian white noise $\b{f}_{\b{Q}}$ is at the same temperature as $\b{f}_{\rho}$ with cross-correlations $\langle\del\cdot\b{f}_{\rho}\b{f}_{\b{Q}}\rangle\propto 2k_BT\nu(\del\del)_{\mathrm{ST}}$ (made symmetric and traceless). Correspondingly the order parameter dynamics is given by $\partial_t\b{Q}=-\delta\mathcal{F}/\delta\b{Q}-\nu(\del\del)_{\mathrm{ST}}\delta\mathcal{F}/\delta\rho+\b{f}_{Q}$, with the Onsager dissipative coefficient $\nu$ included. Apart from relating $K,L_1,L_2$ to two Frank constants, we also have $\lambda=c'_2-\nu c_0/2$ \cite{ostlund1982dynamics}. So, even though the elastic stress $\b{\sigma}^e\simeq\delta\mathcal{F}/\delta\b{Q}$ to lowest order in gradients, generates a term \cite{brochard1977dynamical,ganapathy2007superdiffusion} in $\b{j}_{\mathrm{curl}}=\nu[a_0(\rho_0-\rho_c)-b|\b{Q}|^2/2]\del\cdot\b{Q}$, the coefficient in front being derived from the free-energy $\mathcal{F}$ necessarily vanishes in the ordered phase \footnote{In the isotropic phase, this term is present but doesn't lead to large number fluctuations as $\b{Q}$ is a fast mode.}. 
In that case, the free energy only penalizes gradients of the director, leading to an elastic stress that is $\mathcal{O}(\del^2)$ and hence subdominant in a gradient expansion. The crucial distinction in the active nematic is that violating detailed balance liberates the dynamics from free-energy constraints and the fluctuation-dissipation theorem. The activity $\alpha_0$ has no \emph{apriori} reason to vanish or decrease in correlation with increasing nematic order, with the removal of this constraint being directly responsible for large density fluctuations in the active nematic.

\subsection{Driving of a conserved density field by the Nambu-Goldstone mode}
\label{subsec:thetarho}
Having written down the most general set of equations that governs \emph{any} $2d$ active nematic on a substrate, we now focus on the dynamics deep in the ordered phase. As $\b{Q}$ is symmetric and traceless, in two dimensions it only has two independent components, which we can package into a single complex field $\chi=Q_{xx}+iQ_{xy}$ \footnote{In group theoretic terms, we choose an irreducible complex representation of $\mathsf{U}(1)\times\overline{\mathsf{U}}(1)/\mathbb{Z}_2$ over $\b{Q}$ which transforms as a doublet in the real fundamental tensor representation of $\mathsf{SO}(2)\times\mathsf{SO(2)}/\mathbb{Z}_2$.}. In terms of the angle $\theta$ of the director $\hat{\vec{n}}=(\cos\theta,\sin\theta)$, $\chi=|\chi|e^{2i\theta}$, the factor of $2$ due to the nematic symmetry in the system. As an aside, it is worthwile to note that in $2d$, there is no difference between a polar (vectorial) and apolar (nematic) field (at the level of the equations themselves) as long as one does not mix spatial and field indices. If such a separation is imposed the spatial rotations and rotations of the order parameter field decouple and become independent symmetry operations. The difference in the global structure of the order parameter spaces in the two cases manifests itself only through the character of the topological defects. If (as in our case and as inevitable in a general liquid-crystal system) one does have spatial indices contracted with field indices, then only the combined simultaneous rotation of both spatial coordinates and the order parameter field together becomes a symmetry operation, in which case the terms permitted in the equations themsleves now do depend explicitly on the nature of the field itself. In an active polar fluid, this is manifest by the dual role played by the polar order parameter by also being a velocity that transforms under rotations as the coordinate axes, crucially allowing for the convective nonlinearity that leads to long ranged polar order even in $2d$ \cite{marchetti2013hydrodynamics}. For ease of notation, we shall often switch between $\chi$ as a complex field or a vector like object (its transformation under a rotation is addressed in Sec.~\ref{subsec:symmetry}), the form determined from context. Note, however, that while in a polar fluid the vector order parameter $\chi$ is also a flow velocity, this is not the case in the nematic. In terms of $\chi$, neglecting the elastic anisotropy for the time being, the order parameter equation (Eq.~\ref{eq:Qeq}) becomes
\begin{equation}
\label{eq:chigeneral}
	\partial_t\chi=(a-b|\chi|^2)\chi+K\del^2\chi-\lambda\Gamma\delta\rho+f\ ,
\end{equation}
where $f=f_{\b{Q}xx}+if_{\b{Q}xy}$ is the corresponding noise and $\Gamma=\Gamma_1+i\Gamma_2$ is an anisotropic differential operator ($\Gamma_1=\partial_x^2-\partial_y^2$ and $\Gamma_2=2\partial_x\partial_y$).

Deep in the ordered state, for $\rho=\rho_0$ we have $a=a_0>0$ and $|\chi|\neq 0$. Setting $|\chi|=S_0+\delta S$, where $S_0=\sqrt{a_0/b_0}$ and $\delta S$ is a small fluctuation in the scalar order parameter, we can slave the fast amplitude fluctuations to the remaining slow modes: the phase (being a Nambu-Goldstone mode) and the density (being a conserved field). Neglecting $\partial_t\delta S$ at long time gives
\begin{gather}
	\partial_t\delta S=S_0\left[(a_1-b_1S_0^2)\delta\rho-2b_0S_0\delta S\right]-K S_0|\del\theta|^2\ ,\\
	\implies\delta S\simeq\dfrac{a_1-b_1S_0^2}{2b_0S_0}\delta\rho-\dfrac{K}{2b_0S_0}|\del\theta|^2+\cdots\ ,
\end{gather}
As $a_1=\partial a/\partial\rho$ and $b_1=\partial b/\partial\rho$, both evaluated at $\rho=\rho_0$, the coefficient in front of $\delta\rho$ above can be of either sign and is non-vanishing in general.
Including the elastic anisotropies only leads to anisotropic terms in $\theta$ of the same order as $|\del\theta|^2$. As we shall see, all gradient contributions to $\delta S$ are irrelevant at long distances by power counting. So keeping only the first term, we include the most relevant contribution of the amplitude fluctuation in the equations for the slow modes. Upon doing so, the density equation now takes the form 
\begin{equation}
	\partial_t\delta\rho=D\del^2\delta\rho+\delta D\Gamma\cdot(\hat{\chi}\delta\rho)-\alpha\Gamma\cdot\hat{\chi}-D_n\del\cdot(\hat{\vec{v}}\delta\rho)+\sigma\del^2\delta\rho^2\ .\label{eq:rho}
\end{equation}
Here $\alpha=\alpha_0M_1S_0$ is the lowest order active current contribution, $D=c_0 M_1$ is a regular diffusion constant, $\delta D$ and $D_n$ are anisotropic diffusion constants (in equilibrium $\delta D=D_n$, but active corrections $\propto\alpha_0,\alpha_1$ make them different) and $\sigma=M_1c_1+c_0M'_1/2$ is a passive interaction contribution to the diffusion flux. If we were to write $\b{j}_{\mathrm{curl}}$ as the divergence of an active stress, then $\alpha>0$ would correspond to a contractile system and $\alpha<0$ to an extensile one. We have neglected the conserving noise $\b{f}_{\rho}$ as its effects are subdominant at long wavelengths to those of the orientational noise $\b{f}_{\b{Q}}$ entering via $\alpha\Gamma\cdot\hat{\chi}$. Here $\hat{\chi}=(\cos 2\theta,\sin 2\theta)$ (or $e^{2i\theta}$ in complex form) and $\Gamma=(\partial_x^2-\partial_y^2,2\partial_x\partial_y)$ is the anisotropic differential operator introduced just after Eq. (\ref{eq:chigeneral}). We also have $\vec{v}=\del\cdot\b{Q}$, which in terms of $\chi$ is given by $v_x=\del\cdot\chi$ and $v_y=\del\times\chi$ ($\hat{\vec{v}}$ is similarly given in terms of $\hat{\chi}$). Similarly, the equation for the director phase is given by (upto a rescaling of variables)
\begin{align}
	\partial_t\theta&=K\del^2\theta+\delta K\hat{\chi}\cdot\Gamma\theta+g\del\delta\rho\cdot\del\theta+\kappa\delta\rho\del^2\theta\nonumber\\
					&\quad-\lambda\hat{\chi}\times\Gamma\delta\rho+K_n\hat{\vec{v}}\cdot\del\theta+f_{\theta}\ .\label{eq:theta}
\end{align}
The nonlinear coupling $g$ (depending on $a,b$) arises from amplitude fluctuations of $\delta S$, $K=(K_1+K_3)/2$ is the average Frank elastic constant of the nematic, $\kappa$ is the leading density dependence of the average elastic constant $K$ and $\delta K=L_1S_0=(K_3-K_1)/2$ is the Frank constant anisotropy ($K_1$ and $K_3$ being the splay and bend elastic constants respectively). $K_n=L_2S_0$ is also an independent elastic anisotropy related to $\delta K$ only at equilibrium. The cross coupling $\lambda$ is a consequence of flow alignment, corresponding to the rotation of the nematic director in the presence of a mass flux. As $\b{f}_{\b{Q}}$ is gaussian white noise, the noise in the director phase $f_{\theta}=(\cos2\theta f_{\b{Q}xy}-\sin2\theta f_{\b{Q}xx})/2S$ is also gaussian with a vanishing mean ($\langle f_{\theta}\rangle=0$). The two point correlation is given by $\langle f_{\theta}^2\rangle=(\cos^22\theta\langle f_{\b{Q}xy}^2\rangle+\sin^22\theta\langle f_{\b{Q}xx}^2\rangle-2\sin2\theta\cos2\theta\langle f_{\b{Q}xx}f_{\b{Q}xy}\rangle)/2S$. As both $f_{\b{Q}xx}$ and $f_{\b{Q}xy}$ are independent and identically distributed $\delta$-correlated random variables, the cross terms vanish and we get
\begin{equation}
	\langle f_{\theta}(\vec{r},t)f_{\theta}(\vec{r}',t')\rangle=\Delta\delta(\vec{r}-\vec{r}')\delta(t-t')\ .
\end{equation}
Here we have absorbed factors of two and $S_0$ into the noise variance $\Delta$ and neglected multiplicative noise corrections in $\delta S\sim\delta\rho$.

As we wish to perform a low noise expansion about the ordered state, fluctuations in $\theta$ and $\delta\rho$ are consequently small. Hence the entire analysis is essentially of a ``spin-wave'' type. In equilibrium, both $2d$ polar and nematic liquid crystals (even when compressible) have the same long-distance description as that of the XY model \cite{nelson1977momentum}, in which the spin-wave theory is free and one requires topological defects to proliferate and disorder the system \cite{berezinskii1971destruction,kosterlitz1973ordering,kosterlitz1974critical}. In the active nematic, the Nambu-Goldstone mode interacts with itself due to the nematic anisotropy (as would be in the case of unequal Frank constants \cite{nelson1977momentum}), but also strongly with the density field, in which it engenders large fluctuations. As a consequence, infrared singularities occur in both slow fields, making the question of the stability of the ordered phase rather subtle. What makes the ordered phase of the active nematic so drastically different from its equilibrium counterpart is this invasion of the broken-symmetry mode into the density dynamics.

\section{Linearized Hydrodynamics and the Gaussian Fixed Point}
\label{sec:linhydro}
Starting with an ordered state in the $x$-direction, without loss of generality, the linearized equations for small $\theta$ and $\delta\rho$ are given by
\begin{align}
	\partial_t\delta\rho&=D\del^2\delta\rho+\delta D\Gamma_1\delta\rho-2\alpha\Gamma_2\theta\ ,\label{eq:linrho}\\
	\partial_t\theta&=K\del^2\theta+\delta K\Gamma_1\theta-\lambda\Gamma_2\delta\rho+f_{\theta}\ .\label{eq:lintheta}
\end{align}
There are two primary consequences of activity. The first is seen even at the linear level in the curvature current $\propto\alpha$. The nonlinear effects of this term are addressed in this paper. The second is the motion of defects, i.e. the fact that $+1/2$ disclinations become motile and self-propelled \cite{narayan2007long,giomi2013defect}. This is necessarily non-perturbative and far beyond the scope of the present work, and will be addressed elsewhere.
Fourier transforming $\Phi_{\b{q},\omega}=\int\dd^2r\Phi(\b{r},t)e^{-i\b{q}\cdot\b{r}+i\omega t}$ with $\Phi=(\theta,\delta\rho)$, the inverse propagator for the linearized gaussian theory is given by
\begin{widetext}
	\vspace{-1.5em}
\begin{gather}
	G^{-1}(\b{q},\omega)=\left(\begin{matrix}
			-i\omega+Kq^2+\delta K(q_x^2-q_y^2) & -2\lambda q_xq_y\\
			-4\alpha q_xq_y & -i\omega+Dq^2+\delta D(q_x^2-q_y^2)
	\end{matrix}\right)\ ,
\end{gather}
\end{widetext}
where $q=|\b{q}|$. The detailed angular dependence of the eigenmodes is given in Ref.~\cite{ramaswamy2003active}.

We require $K,D>0$, $|\delta K|< K,|\delta D|< D$ and $\alpha\lambda$ not be too large for stability (for $\delta D=\delta K=0$, the stability line is given by $\alpha\lambda<KD/2$, the general criterion being more involved). These stability lines correspond to splay-bend instabilities that have a finite threshold due to the presence of a frictional substrate and have been extensively studied (see for instance Refs.~\cite{putzig2016instabilities,srivastava2016negative} and reference therein), so we shall not discuss them any further. Note that, as we are deep in the ordered phase, we do not concern ourselves with the density banding instability which only occurs near the mean-field transition.

Within the gaussian theory, we can easily compute the density and angle correlators. For simplicity, we shall consider $\delta K,\delta D$ and $\alpha\lambda\ll D,K$, in which case
\begin{gather}
	\langle\left|\theta_{\b{q},\omega}\right|^2\rangle\approx\dfrac{\Delta}{\omega^2+K^2q^4}\ ,\\
	\langle\left|\delta\rho_{\b{q},\omega}\right|^2\rangle\approx\dfrac{16\Delta \alpha^2q_x^2q_y^2}{(\omega^2+D^2q^4)(\omega^2+K^2q^4)}\ .
\end{gather}
Going back to real space, the equal time two-point correlator of the $2p$-atic order parameter $\Psi_{2p}=e^{2ip\theta}$ (for $p=1$, $\Psi_2=\hat{\chi}$, the unit normalized complex nematic order parameter we had before) is given by
\begin{equation}
	\langle\Psi_{2p}^{\ast}(\vec{r},t)\Psi_{2p}(0,t)\rangle=\left(\dfrac{r}{a}\right)^{-\eta_p(\Delta)}\ ,
\end{equation}
where $a$ is some microscopic cutoff and $\eta_p(\Delta)=p^2\Delta/2\pi K$ is a non-universal exponent (as it depends on the strength of the noise and the elastic stiffness), that governs the power-law decay of the order parameter. So, the linearized equations only predict quasi-long-ranged order (QLRO), just like in equilibrium ($\langle\Psi_{2p}\rangle=0$ in the thermodynamic limit).

Though the active current, at the linear level so far, does not alter the conclusion of quasi-long-range order in $2d$, it does leave a rather spectacular footprint on the density fluctuation spectrum, which was shown \cite{ramaswamy2003active} to diverge as $\b{q}\rightarrow0$. The equal time structure factor is given by
\begin{align}
	S(\b{q})&=\langle\delta\rho_{\b{q}}(t)\delta\rho_{-\b{q}}(t)\rangle\nonumber\\
					 &=\dfrac{8\Delta \alpha^2}{DK(D+K)}\dfrac{q_x^2q_y^2}{q^6}\sim\dfrac{1}{q^2}\quad\mathrm{as}\ \b{q}\rightarrow0\label{eq:Sqlin}\ .
\end{align}
As the number fluctuations in a volume $V\sim L^2$ scale as $\sqrt{\langle\delta N^2\rangle}\sim\sqrt{S(\b{q}\rightarrow0)V}$, this gives $\sqrt{\langle\delta N^2\rangle}\propto N$ \cite{ramaswamy2003active}. Later in Sec.~\ref{sec:drg}, we shall show how nonlinearities modify this result and change the GNF exponent to a non-universal number.

Note that even though we do not have long-ranged orientational order, the structure factor in Eq.~\ref{eq:Sqlin} is markedly anisotropic. This is an artifact of having performed a linearization around the $x$-axis $\theta=0$ state. Fixing a global frame of reference, the above result is an average within a restricted ensemble of a fixed refernce state. Linearizing about a reference state at $\theta=\theta_0$, we instead obtain
\begin{equation}
	S(\b{q}\ ;\theta_0)=\dfrac{2\Delta\alpha^2}{K D(K+D)q^6}[2\cos2\theta_0 q_xq_y-\sin2\theta_0(q_x^2-q_y^2)]^2\ .
\end{equation}
The absence of long-ranged order means that the steady state distribution of the reference angle is uniform over the $[0,\pi)$ interval. Using $P(\theta_0)=1/\pi$, we average $S(\b{q}\ ;\theta_0)$ over $\theta_0$ to correctly recover isotropy in the density correlator,
\begin{equation}
	S(\b{q})=\overline{S(\b{q}\ ;\theta_0)}=\dfrac{\Delta\alpha^2}{KD(K+D)}\dfrac{1}{q^2}\ .\label{eq:Sqiso}
\end{equation}

In order to assess the importance of the nonlinearities, we perform the following scalings
\begin{gather}
	t\rightarrow b^z t\ ,\quad\vec{r}\rightarrow b\vec{r}\ ,\\
	\delta\rho\rightarrow b^{\zeta_{\rho}}\delta\rho\ ,\quad\theta\rightarrow b^{\zeta_{\theta}}\theta\ ,
\end{gather}
where $z$ is a dynamical exponent and $\zeta_{\rho}$ and $\zeta_{\theta}$ are ``roughness'' exponents for the two fields. This gives the following scaling dimensions
\begin{gather}
	y_{\Delta}=z-2\zeta_{\theta}-2\ ,\\
	y_{D}=y_{K}=z-2\ ,\quad y_{\delta K}=y_{\delta D}=z-2\ ,\\
	y_{\alpha}=z-2+\zeta_{\theta}-\zeta_{\rho}\ ,\quad y_{\lambda}=z-2+\zeta_{\rho}-\zeta_{\theta}\ .
\end{gather}
Hence at the linear fixed point, requiring that all the linear terms and the noise variance $\Delta$ not change under this scaling fixes
\begin{align}
	z=2\quad\mathrm{and}\quad\zeta_{\rho}=\zeta_{\theta}=\dfrac{z-2}{2}=0\ .
\end{align}
Above two dimensions, the linearized description is correct with all the non-linearities being irrelevant, but in exactly two dimensions both the fields $\theta$ and $\delta\rho$ become marginal and dimensionless. As the scale of density fluctuations is the same as that in the phase, setting $\zeta_{\rho}=\zeta_{\theta}=\zeta$, a nonlinear term of the kind
\begin{equation}
	\del^{2+2k}\delta\rho^{n}\theta^{m}\sim b^{z-2-2k+(n+m-1)\zeta}\ ,
\end{equation}
present in either equation is \emph{marginal} in the infrared for $k=0$ and \emph{all} $n,m>1$ for $d=2$ dimensions. Higher gradient terms with $k>1$ are infrared \emph{irrelevant} by simple power counting at the gaussian fixed point. Hence, the two dimensional active nematic has an \emph{infinite} spectrum of marginal operators at the linear fixed point, much like the situation for regular three dimensional Navier-Stokes turbulence \cite{eyink1994renormalization}. In order to judge the (un)importance of any of the marginal nonlinearities, one is immediately forced to take recourse to a dynamical renormalization group programme, but the fact that an infinity of them have to be handled seems unsurmountable. This is where the symmetries of this system provide a great simplification.

\subsection{Rotations and symmetries}
\label{subsec:symmetry}
The true symmetry of a nematic liquid crystal with unequal Frank elastic constants is one in which both spatial coordinates and the director field are rotated by the same angle. In two dimensions, a rotation by an angle $\varphi$ is given by the following matrix
\begin{equation}
	R(t)=\left(
	\begin{matrix}
		\cos\varphi & -\sin\varphi\\
		\sin\varphi & \cos\varphi
	\end{matrix}
	\right)\ .
\end{equation}
Hence, the symmetry transformation is then given by $\vec{x}\rightarrow\vec{x}'=R\vec{x}$, $\b{Q}\rightarrow\b{Q}'=R\b{Q}R^{T}$ ($\theta\rightarrow\theta'=\theta+\varphi$, where $\theta$ is the angle of the director and $\varphi$ the rotation angle). For an infinitesimal rotation by $\epsilon$, the derivatives transform as $\partial_x\rightarrow\partial_x'=\partial_x-\epsilon\partial_y$ and $\partial_y\rightarrow\partial_y'=\partial_y+\epsilon\partial_x$. This in turn leads to the following transformations for the anisotropic differential operator $\Gamma$.
\begin{align}
	\Gamma_1=\partial_x^2-\partial_y^2&\rightarrow\Gamma_1'=\Gamma_1-2\epsilon\Gamma_2\ ,\\
	\Gamma_2=2\partial_x\partial_y&\rightarrow\Gamma_2'=\Gamma_2+2\epsilon\Gamma_1\ .
\end{align}
As $\hat{\chi}=(\cos2\theta,\sin2\theta)$ also transforms with $\theta\rightarrow\theta'=\theta+\epsilon$, we immediately note that $\hat{\chi}\cdot\Gamma$ and $\hat{\chi}\times\Gamma$ are invariant under this symmetry operation (along with the obvious isotropic laplacian $\del^2$). Additionally, we have $\vec{v}=\del\cdot\b{Q}=(\del\cdot\chi,\del\times\chi)$, which transforms as a vector. So including $\vec{v}\cdot\del$ and $|\vec{v}|^2$ ($\del\cdot\vec{v}=\Gamma\cdot\chi$), we exhaust all the scalar terms that are allowed by rotational symmetry.

Apart from being apolar, a nematic liquid crystal is also \emph{achiral}. Specifically, choosing a local orthogonal frame $\{\partial_x,\partial_y\}$ with the $x$-axis aligned along the local orientation, in the absence of local enantiomorphy or molecular chirality, we also require the invariance under local parity reflections: $y\rightarrow-y$ (if the frame weren't oriented along the local director orientation, then additionally one must also flip the director angle $\theta\rightarrow-\theta$) \cite{frank1958liquid}. Under this action $\Gamma_1\rightarrow\Gamma_1$ and $\Gamma_2\rightarrow-\Gamma_2$, which immediately shows that $\chi\cdot\Gamma$ is even under reflections while $\chi\times\Gamma$ is odd. Hence we must additionally only include terms in the equation that preserve the parity of the variables ($\rho$ being parity even and the phase $\theta$ parity odd).

Expanding in small fluctuations of $\theta,\delta\rho$, these symmetries provide powerful constraints on the possible nonlinear mode-coupling terms that can be present. In particular, the full rotational symmetry of the model is nonlinearly realized in the broken symmetry mode $\theta$, so one must treat all terms related by a symmetry transformation on an equal footing. As we show in Sec.~\ref{sec:drg} and the Appendix~\ref{sec:aniso}, all the terms that are explicitly anisotropic (linear or nonlinear) are marginally irrelevant at leading order just as a consequence of rotational symmetry. This allows us to directly disregard most of the nonlinear couplings involving $\b{Q}$ that one would write down. Hence only a small fraction of these anisotropic terms have to be considered, with the most important nonlinearities arising from expanding mode-coupling terms in Eqs.~\ref{eq:rho} and~\ref{eq:theta} that also contribute at the linear level. For example the active current term $\alpha\Gamma\cdot\hat{\chi}\simeq2\alpha\Gamma_2\theta-2\alpha\Gamma_1\theta^2$ ($\theta\ll1$) is present in the linear equations (Eq.~\ref{eq:linrho}) and also generates a nonlinear interaction term $-2\Gamma_1\theta^2$ among others, with the exact same coefficient $\alpha$. Such relations being a consequence of symmetry must be preserved under renormalization. A well known example of singular fluctuation corrections arising from symmetry-required nonlinearities is the elasticity and hydrodynamics of an equilibrium smectic liquid crystal \cite{grinstein1981anharmonic,grinstein1982nonlinear,mazenko1982viscosities,mazenko1983breakdown,milner1986fluctuating}. In addition to a plethora of anisotropic terms, there are also isotropic nonlinearities one has to keep track of, for example the terms $\sigma\del^2\delta\rho^2$, $\kappa\delta\rho\del^2\theta$ and $g\del\theta\cdot\del\delta\rho$ in Eqs.~\ref{eq:rho} and~\ref{eq:theta} respectively. These terms come with independent coefficients unrelated to any other coupling constants and don't affect the linear hydrodynamic description.

So anticipating ourselves, we neglect all higher order anisotropic nonlinearities (like $K_n,D_n$), while only retaining the symmetry required and isotropic ones, the assumption of irrelevance being justified \emph{a posteriori}. Keeping this in mind, the full set of dynamical equations for small fluctuations in $\delta\rho$ and $\theta$ is given by
\begin{widetext}
	\vspace{-1.5em}
\begin{align}
	\partial_t\delta\rho&=D\del^2\delta\rho+\delta D\left[\Gamma_1\delta\rho+\Gamma_2(2\theta\delta\rho)-\Gamma_1(2\theta^2\delta\rho)\right]-\alpha\left[\Gamma_2(2\theta)-\Gamma_1(2\theta^2)-\dfrac{4}{3}\Gamma_2(\theta^3)\right]+\sigma\del^2\delta\rho^2\ ,\label{eq:rhoeqn}\\
	\partial_t\theta&=K\del^2\theta+\delta K\left[\Gamma_1\theta+2\theta\Gamma_2\theta-2\theta^2\Gamma_1\theta\right]+g\del\delta\rho\cdot\del\theta+\kappa\delta\rho\del^2\theta-\lambda\left[\Gamma_2\delta\rho-2\theta\Gamma_1\delta\rho-2\theta^2\Gamma_2\delta\rho\right]+f_{\theta}\ .\label{eq:thetaeqn}
\end{align}
\end{widetext}

\section{Perturbative Dynamical Renormalization}
\label{sec:drg}
Following \citet{forster1977large}, we perform a 1-loop computation of the renormalization flow equations perturbatively in the nonlinearities. As each loop correction comes with an accompanying factor of the noise variance $\Delta$, the loop expansion corresponds precisely to a systematic and controlled low noise expansion. Fixing an ultraviolet cutoff in fourier space $|\b{q}|<\Lambda(=1/a)$, we split the fields into slow and fast modes ($\theta=\theta_{<}+\theta_>$, $\delta\rho=\delta\rho_<+\delta\rho_>$) and coarse-grain out short scale fluctuations in a momentum shell $\Lambda/b<|\b{q}_{>}|<\Lambda$, which after appropriate rescaling of the coordinates and the fields gives an equation of the same form as we have written above (Eqs.~\ref{eq:rhoeqn},~\ref{eq:thetaeqn}), though now with modified coefficients. Finally letting $\ln b=\ell\ll1$, we obtain differential flow equations that govern the long-wavelength behaviour of the theory as we iterate the coarse-graining procedure out to the largest scales of interest. The renormalized propagator $G_{R}(\b{q},\omega)$ satisfies the following Dyson equation
\begin{equation}
	G_{R}^{-1}(\b{q},\omega)=G^{-1}(\b{q},\omega)-\Sigma(\b{q},\omega)\ ,
\end{equation}
where $\Sigma(\b{q},\omega)$ is the ``self-energy'' that includes all the diagrammatic contributions. The details of this long set of computations is given in Appendix~\ref{sec:renorm}, and we shall only briefly sketch and analyze the main results here.

For small $\Delta$, using the result of the linearized analysis, one can obtain the leading fluctuation corrected linear theory due to the interaction with the Nambu-Goldstone mode $\theta$. We shall illustrate this here for the diffusive anisotropy $\delta D$. Considering $\alpha,\lambda,\delta K,\delta D,g,\kappa$ and $\sigma$ to all be sufficiently small, to leading order the joint probability distribution of $\delta\rho$ and $\theta$ essentially factors (as the cross couplings $\alpha$ and $\lambda$ are small). In this limit, corrections to $K,D$ and $\Delta$ are negligible and we can estimate the effect phase fluctuations have on the anisotropic terms. Averaging just over $\theta$, for the diffusive anisotropy, we have
\begin{equation}
	\delta D\left[\Gamma_1(\delta\rho\cos2\theta)+\Gamma_2(\delta\rho\sin2\theta)\right]\approx\delta D\langle\cos2\theta\rangle\Gamma_1\delta\rho\ ,\\
\end{equation}
where we have used the linear theory result $\langle\sin2\theta\rangle=0$ and $\langle\cos2\theta\rangle=(L/a)^{-\eta(\Delta)}$ ($\eta(\Delta)=\Delta/2\pi K$) is a system size $L$ dependent constant, leading to a renormalized diffusion anisotropy $\delta D(L)=(L/a)^{-\eta(\Delta)}\delta D$. This immediately tells us that the fluctuations of the director phase cause anisotropic terms such as the one above to become length scale dependent, driving them to zero as a power law in larger and larger systems. In Appendix~\ref{sec:aniso}, we systematically show that this leading behaviour is a consequence of rotational symmetry of the model and is hence true for \emph{all} anisotropic terms, be they linear or nonlinear (i.e. any term involving a contraction of order parameter and spatial indices). This point is crucial as it allows us to immediately treat an infinite number of anisotropic nonlinearities, showing them to be marginally irrelevant at least at leading order and justifies their neglect in Eqs.~\ref{eq:rhoeqn} and \ref{eq:thetaeqn}. Hence the only important \emph{anisotropic} terms will have to be the ones present in the linearized equations. This is precisely why we only kept those nonlinearities that are related by symmetry to linear terms and disregarded all the higher order anisotropies (like $K_n,D_n$). Note that this argument does not work for the \emph{isotropic} nonlinearities, which still do need to be treated by the full renormalization group analysis.

At this order, as $K,D$ and $\Delta$ remain unrenormalized and don't run with scale, we see that the orientational order remains quasi-long ranged, but the density fluctuations now become anomalous. As the active current involves a contraction of order parameter and spatial indices, $\alpha$ is also an anisotropic coupling and it runs with scale in the same fashion as above - $\alpha(\b{q})\sim q^{\eta(\Delta)}$ having switched to a wavevector representation. Using this renormalized activity, as $\b{q}\to 0$, we find
\begin{equation}
	\langle|\delta\rho_{\b{q}}(t)|^2\rangle\sim\dfrac{\Delta\alpha(\b{q})^2q_x^2q_y^2}{KD(K+D)q^6}\propto L^{2-2\eta(\Delta)}\ ,\label{eq:Sq}
\end{equation}
with $2\pi/q=L$ as the longest wavelength in a system of linear size $L$. The anisotropy here is still a consequence of a restricted ensemble average and not of long-ranged order. A complete ensemble average recovers isotropy in the density correlator as discussed before (see Eq.~\ref{eq:Sqiso}).

Even though the strength of the active drive gets renormalized to zero, one \emph{does not} recover an equilibrium system. Hence activity is \emph{dangerously} irrelevant \cite{amit1982dangerous}, leaving a strong imprint on the fluctuations even as it vanishes at large scales. This is similar in spirit to dangerously irrelevant hexagonal symmetry breaking perturbations controlling the divergence of the longitudinal susceptibility in an ordered ferromagnet \cite{nelson1976coexistence}. Here instead, the active drive is a (marginally) irrelevant \emph{detailed-balance-breaking} perturbation and its consequences remain non-negligible even for asymptotically small activity. With this modification to the structure factor, we find that the giant number fluctuations continue to persist, but with a modified non-universal scaling exponent, suppressed from its linearized prediction by a noise dependent number $\eta(\Delta)$,
\begin{equation}
	\sqrt{\langle\delta N^2\rangle}\propto N^{1-\eta(\Delta)/2}\ .\label{eq:gnf}
\end{equation}
An informal shortcut to this result is to note that the active current $\sim \del_i\del_jQ_{ij}$ in the density equation is proportional to the nematic order parameter $\b{Q}$. Grafting this onto the linearized calculation \cite{ramaswamy2003active} of Sections \ref{subsec:thetarho} and \ref{sec:linhydro} shows that GNFs are mitigated by the same power of system size as the order parameter, i.e., the QLRO exponent $\eta(\Delta)$. This directly gives an improved estimate for the density fluctuation variance as $\propto L^{2-2\eta(\Delta)}$, which is the same as that given above in Eq.~\ref{eq:Sq}.


As the density fluctuation $\delta\rho$ is a conserved variable, all the nonlinear interaction terms have to be the divergence of some current. From Eq.~\ref{eq:rhoeqn}, if we set $\sigma=0$ we see that nonlinear terms involving either $\alpha$ or $\delta D$ are anisotropic total derivatives and hence give rise to only anisotropic corrections in $\Sigma_{\rho\rho}\propto q_x^2-q_y^2$, thereby leaving the isotropic diffusion constant unrenormalized to all orders in perturbation theory. For $\sigma\neq0$, the most relevant contribution to the diffusion propagator is $\Sigma_{\rho\rho}\sim\Delta\sigma\alpha\delta D q^2$, which corrects $D$ by a small amount (including $D_n\neq0$, there are small $\mathcal{O}(\Delta D_n^2)$ corrections as well which are irrelevant as $D_n$ itself is irrelevant). As we assume all the couplings and noise (except for $K,D$) to be small, this correction is already far smaller than the leading corrections we shall be interested in (noise times two coupling constants). So to this level of approximation within perturbation theory, the diffusion constant $D$ is nearly unrenormalized, hence,
\begin{equation}
	\dfrac{\dd D}{\dd\ell}=D(z-2)\ .
\end{equation}
Fixing the dynamical exponent $z=2$, we can keep $D$ fixed at its bare microscopic value, which we set to unity ($D=1$) from now on, without loss of generality.

Given the large number of parameters, for the purposes of this discussion we restrict ourselves to the case of vanishing elastic and diffusive anisotropies ($\delta K=\delta D=0$) and $g=\kappa=0$. This corresponds to an invariant subspace of the flow equations given in Appendix~\ref{sec:renorm}. This is sufficient to elucidate the main consequences of activity at the nonlinear level, as this surface is stable and attracting, with small deviations from it being irrelevant (see Appendix~\ref{sec:renorm} for more details). This simplification decouples most of the flow equations, leaving us with only two coupled ones.
\begin{figure}[t]
	\centering
	\includegraphics[width=0.47\textwidth]{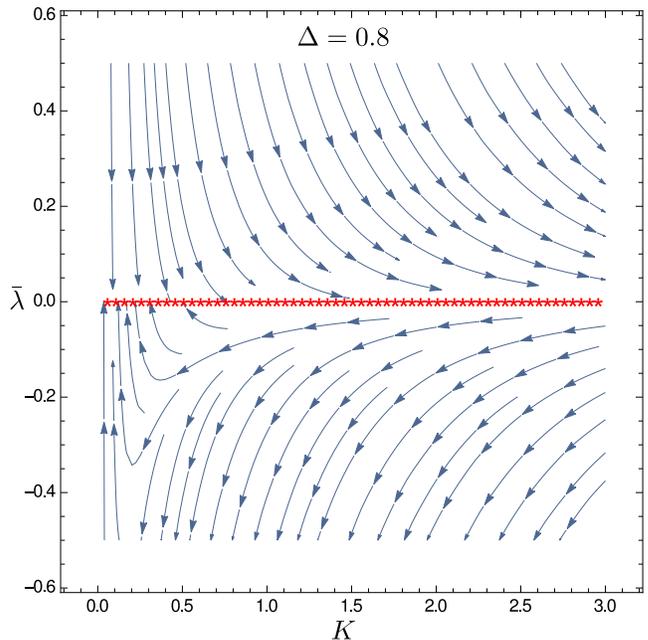}
	\caption{The low noise renormalization group flow diagram in the case of vanishing anisotropy ($\delta D=\delta K=0$) and $g=\kappa=0$. The noise variance is fixed at $\Delta=0.8$. The red stars at $\bar{\lambda}=0$ correspond to a line of fixed points in $K$. For a large elastic stiffness $K$, and $\bar{\lambda}>0$, it takes a large number of renormalization group iterations in order to reach the fixed point, with a relatively large intermediary regime in which the elastic constant grows logarithmically with scale (Eq.~\ref{eq:Kpowerlaw}). For most $\bar{\lambda}<0$, the flow is unstable with decreasing $K$, possibly going to a strong coupling fixed point not accessible within perturbation theory.}
	\label{fig:RGflow}
\end{figure}
\begin{align}
	\dfrac{\dd K}{\dd\ell}&=Kc_K\bar{\lambda}\ ,\label{eq:Kflow}\\
	\dfrac{\dd\bar{\lambda}}{\dd\ell}&=-\bar{\lambda}\left(\dfrac{\Delta}{\pi K}+b_K\bar{\lambda}\right)\ ,\label{eq:lambdabarflow}
\end{align}
where $\bar{\lambda}=\alpha\lambda\Delta/[\pi K^2(1+K)]$ is a non-dimensional active coupling. In this limit of $\delta K=\delta D=g=\kappa=0$, the noise variance $\Delta$ also remains unrenormalized at leading order, fixed at its microscopic value. Both $c_K$ and $b_K$ are positive, monotonic functions of $K$ that remain finite in both the limits $K\to0$ and $K\to\infty$,
\begin{equation}
	c_K=\dfrac{1+3K+4K^2}{(1+K)^2}\ ,\quad b_K=1+c_K\left(\dfrac{2+3K}{1+K}\right)\ .
\end{equation}
As $K\to 0$, $c_K\sim1$ and $b_K\sim 3$, while as $K\to\infty$, $c_K\sim 4$ and $b_K\sim 13$. The renormalization group flow diagram within the $\{K,\bar{\lambda}\}$ subspace for a fixed $\Delta$ is shown in Fig.~\ref{fig:RGflow}. At a given noise variance $\Delta$, for low enough activity, we can neglect the second $\bar{\lambda}^2$ term in Eq.~\ref{eq:lambdabarflow}. Treating $c_K$ to be essentially constant, we can then integrate the flow equations approximately to get,
\begin{equation}
	\dfrac{\dd K}{\dd\bar{\lambda}}=-\dfrac{\pi c_K}{\Delta}K^2\implies\dfrac{1}{K(\ell)}-\dfrac{1}{K_0}=\dfrac{\pi c_K}{\Delta}\left(\bar{\lambda}(\ell)-\bar{\lambda}_0\right)\ .
\end{equation}
Setting $\ell=\ln(\Lambda/q)$ with $\Lambda=1/a$ being the ultraviolet cutoff, as $\b{q}\to 0$, we have
\begin{equation}
	K_{\infty}-K(\b{q})\propto\left(\dfrac{q}{\Lambda}\right)^{x(\Delta)}\quad\mathrm{and}\quad\bar{\lambda}(\b{q})\propto\left(\dfrac{q}{\Lambda}\right)^{x(\Delta)}\ ,\label{eq:Ksoln}
\end{equation}
where $K_0,\bar{\lambda}_0$ are the microscopic parameters we begin with at short scales, $K_{\infty}=K_0\Delta/(\Delta-\pi c_K K_0\bar{\lambda}_0)$ is the final asymptotic nematic stiffness and $x(\Delta)=\Delta/\pi K_0-c_K\bar{\lambda}_0$ is a non-universal exponent. This solution is only valid as long as $x(\Delta)>0$. The sign of $\bar{\lambda}_0$ depends on both the microscopic activity $\alpha$ and the flow-alignment like parameter $\lambda$. If we suppose $\lambda>0$ for elongated nematogens, then the renormalized elastic stiffness $K_{\infty}>K_0$ for a contractile system and $K_{\infty}<K_0$ for an extensile system. Coarse-graining microscopic models of an active nematic \cite{bertin2013mesoscopic,shi2014instabilities}, or a self-propelled rod system \cite{peshkov2012nonlinear,baskaran2008hydrodynamics}, where the notion of contractile or extensile stresses may not be so obvious, though always give $\bar{\lambda}>0$ leading to a stiffer system at large scales. As $K_{\infty}$ is still finite in the thermodynamic limit we end up only with \emph{quasi-long ranged nematic order}. Once again as we saw earlier, the active coupling is irrelevant at large scales, but dangerously so as its effects on the density fluctuations do not consequently vanish. For $\bar{\lambda}_0<0$, from Fig.~\ref{fig:RGflow}, we see that there is a small region close to the $\bar{\lambda}=0$ line of fixed points where noise nonlinearly stabilizes the system, but elsewhere, $K$ decreases continuously, possibly vanishing or even going negative at some strong coupling fixed point. This would signal a modulational instability, possibly giving rise to a smectic array of bend-splay distortions, about which one would have to reorganize the low noise fluctuation expansion, far beyond the scope of this paper. Note that unlike the linear Lifshitz instability prediction for an overdamped $2d$ active nematic without a conserved density at the mean field level \cite{srivastava2016negative}, here the theory is linearly stable to begin with and only destabilized nonlinearly in the presence of noise.

\begin{figure}[]
	\centering
	\includegraphics[width=0.5\textwidth]{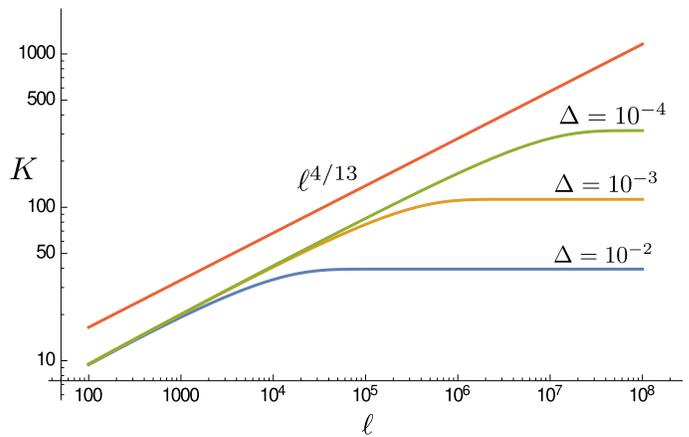}
	\caption{Comparing the growth of $K(\ell)$ for different values of the noise $\Delta$. The initial parameter values are $K_0=1$ and $\bar{\lambda}_0=1$. The red line ($\ell^{4/13}$) is just given as a guideline. Note that $\ell=\ln(L/a)$ at the scale of the system size, it only changes by $\mathcal{O}(1)$ under finite size scaling. The large logarithmic scale shown here in $\ell$ is just to compare the approximate analytical prediction to the numerical solution of the differential equations.}
	\label{fig:Kpowerlaw}
\end{figure}

For larger values of the active drive with $\bar{\lambda}_0>0$, $x(\Delta)$ can be negative and the $b_K\bar{\lambda}^2$ nonlinearity in Eq.~\ref{eq:lambdabarflow} becomes important. As $\Delta<\pi c_K K_0\bar{\lambda}_0$, taking $b_K$ to be nearly constant, we have approximately
\begin{equation}
	\dfrac{\dd\bar{\lambda}}{\dd\ell}\simeq-b_K\bar{\lambda}^2\implies\bar{\lambda}(\ell)\propto\dfrac{1}{b_K\ell}\quad(\mathrm{as}\ \ell\to\infty)\ .
\end{equation}
For $\bar{\lambda}>0$, $K$ increases slowly with scale. Replacing $b_K\sim13$ and $c_K\sim4$ for large $K$, we find the growth of the elastic stiffness to be
\begin{equation}
	\dfrac{\dd\ln K}{\dd\ell}\simeq 4\bar{\lambda}=\dfrac{4}{13\ell}\implies K(\ell)\propto\ell^{4/13}\quad(\mathrm{as}\ \ell\to\infty)\ .\label{eq:Kpowerlaw}
\end{equation}
With $\ell=\ln(\Lambda/q)$, the Frank elastic constant grows logarithmically slowly $K(\b{q})\propto \ln(\Lambda/q)^{4/13}$ for $q\ll\Lambda$. This logarithmic breakdown of hydrodynamics is typical when nonlinearities are marginal by power counting, as is also similarly encountered in $2d$ thermal fluids \cite{forster1977large}, solids and hexatic liquid crystals \cite{zippelius1980large} in equilibrium. Here, though, in the thermodynamic limit the slow growth of $K(\b{q})$ is actually arrested as it saturates at a large but \emph{finite} value. This is because in Eq.~\ref{eq:lambdabarflow}, $b_K\bar{\lambda}\ll\Delta/\pi K$ eventually beyond an exponentially large crossover length scale $\xi_{\ast}\sim a \exp[\bar{\lambda}_0^{4/9}(\pi K_0/\Delta)^{13/9}]$, above which one recovers the kind of behaviour shown in Eq.~\ref{eq:Ksoln}, only now with $K_0$ and $\bar{\lambda}_0$ now evaluated at $\ell_{\ast}=\ln(\xi_{\ast}/a)$. Numerically integrating the flow Eqs.~\ref{eq:Kflow} and~\ref{eq:lambdabarflow}, we find the same behaviour described above for sufficiently low noise, as shown in Fig.~\ref{fig:Kpowerlaw}.

Hence when the active drive is stronger than the noise variance for $\bar{\lambda}>0$, there is a possibly large range of system sizes with $L<\xi_{\ast}$ where one would \emph{not} see conventional quasi-long ranged order with $\langle\Psi_{2p}(\b{r},t)\rangle=(L/a)^{-\eta_p(\Delta)}$. Instead, as
\begin{align}
	\langle\theta(\vec{r},t)^2\rangle&=\int\dfrac{\dd^2q}{(2\pi)^2}\dfrac{\Delta}{2K(q)q^2}\nonumber\\
									 &=\dfrac{\Delta}{36\pi K_0\bar{\lambda}_0}\left[13\bar{\lambda}_0\ln\left(\dfrac{L}{a}\right)\right]^{9/13}\ ,\label{eq:theta2}
\end{align}
the order parameter decreases as a \emph{stretched exponential of the logarithm} of the length scale $L$ as long as $a\ll L<\xi_{\ast}$. Though $\langle\theta^2\rangle$ is not finite implying the absence of true long-ranged order, the decay $\langle\Psi_{2p}(\b{r},t)\rangle=\exp[-p^2\Delta(13\ln(L/a))^{9/13}/9\pi K_0\bar{\lambda}_0^{4/13}]$, slower than any power of $L$, might be mistaken in small systems to be indicative of long ranged order. Our analysis however shows that the $2d$ active nematic is always quasi-long ranged ordered in the thermodynamic limit.

One can also provide general arguments to show that the $2d$ active nematic can only truly support quasi-long ranged order in the asymptotic limit. In crucial distinction from the active polar flocking case, where the convective nonlinearity is \emph{relevant} in $d=2$ \cite{toner2012reanalysis,toner1995long}, all the nonlinear terms in the active nematic model are only marginal in two dimensions (be they active or equilibrium in origin). Marginal terms can only produce logarithmic corrections to scaling simply because they are dimensionless to begin with, leading to the renormalization recursion relations not having a linear term in the coupling. Additionally, in the absence of density fluctuations, we know from the result of Ref.~\cite{mishra2010dynamic}, that we recover an equilibrium XY like description at long distances, which is not surprising as the only nonlinear terms present are anisotropic terms that we have shown to be generally irrelevant to leading order as a consequence of rotational symmetry of the model. Including the active currents coupling to the conserved density field, the new non-equilibrium terms are once again irrelevant to leading order, being anisotropic in nature ($\sim\del\del:\b{Q}$). Note that all the (possibly worrisome) isotropic nonlinearities are present even at equilibrium and cannot conspire by themselves to give rise to long-ranged order, for if that were the case, upon taking the equilibrium limit, the same mechanism must continue to work violating the Mermin-Wagner theorem \cite{mermin1966absence}. This is true even upon including multiplicative noise. So the only way the nonlinearities might give rise to long-ranged order is by mixing with operators that violate detailed balance (coming from activity), but every such active term being anisotropic is irrelevant. Hence all anisotropies and nonlinearities being marginal and irrelevant to leading order, the $2d$ active nematic is always doomed to have a finite elastic stiffness in the thermodynamic limit, without any singular corrections, leading inevitably to only quasi-long ranged order. So activity is only dangerously irrelevant with regard to density fluctuations but doesn't affect the phase fluctuations much, except for inducing strong finite size effects as discussed above. As the ordered nematic phase of a self-propelled rod system also has only two slow modes ($\delta\rho$ and $\theta$), with the velocity always decaying on a finite time scale, the long distance hydrodynamic description of such a phase is identical to the one discussed here. One would have to verify if the long-ranged order claimed in such systems \cite{PhysRevE.95.020601,ginelli2010large} is actually a finite size effect in the sense of Eq.~\ref{eq:theta2}, as the phase fluctuations though not finite, grow slower than a logarithm below the crossover scale $\xi_{\ast}$, with only much larger systems eventually recovering true QLRO.

\section{GNF versus Phase Separation}
\label{sec:phasesep}

It is essential to distinguish giant number fluctuations from phase separation which also trivially exhibits $\langle\delta N^2\rangle\propto N^2$ behaviour due to the formation of clusters in a disordered gas. The linear hydrodynamic treatment of the active nematic also predicts number fluctuations proportional to the mean. The question thus arose whether this was phase separation even deep in the ordered phase \cite{mishra2006active}. Note that a possible phase separated phase in the ordered state is distinct from the inhomogeneous chaotic phase present close to transition which has density bands and clusters, but is orientationally \emph{disordered}. It was suggested in Ref.~\cite{mishra2006active} that the giant number fluctuations in the ordered nematic phase realize a peculiar and delicate form of phase separation, where, instead of forming a single macroscopic dense liquid cluster in a gas, the system perpetually transitions amongst many configurations with a finite number of macroscopic clusters. This phenomenon, christened fluctuation-dominated phase ordering, is ubiquitous in models involving particles sliding on randomly fluctuating surfaces \cite{das2000particles}, where a particle current $\propto\del h$ ($h$ being the height of the surface) drives clustering even in the absence of attractive interactions. The question was investigated only in the context of advection of tracer particles by active directed motion due to orientational curvature \cite{mishra2006active}.

Our results provide the first analytical calculation at the nonlinear level that can address and disentangle these phenomena. In Ref.~\cite{dey2012spatial}, the relation between the structure factor and the scaling of number fluctuations is addressed numerically in detail. The constraints imposed by rotational symmetry of the model force the scaling of the giant number fluctuations to be modified from the linearized prediction,
\begin{equation}
	\sqrt{\langle\delta N^2\rangle}\propto N^{1-\eta(\Delta)/2}\ ,
\end{equation}
for sufficiently small activity compared to the noise. If the active drive is stronger than the noise ($\bar{\lambda}\gg\Delta/K$), then using the flow equations (Eq.~\ref{eq:Kflow}, \ref{eq:lambdabarflow}), we find
\begin{align}
	\sqrt{\langle\delta N^2\rangle}\propto\begin{cases}
		N\left(\ln N\right)^{-5/13},\quad L_{N}<\xi_{\ast}\\
		N^{1-\eta(\Delta)/2},\quad L_{N}>\xi_{\ast}
	\end{cases}\ ,
\end{align}
where $L_N$ is the linear size of a region containing $N$ particles on average. So the number fluctuations are still ``giant'', but for sufficiently large averaging volumes they are always parametrically smaller than the linear prediction. The corresponding angle averaged structure factor looks like
\begin{equation}
	\langle\delta\rho(\b{r},t)\delta\rho(\b{0},t)\rangle\sim\dfrac{1}{r^{2\eta(\Delta)}}\ ,\quad r\to\infty\ ,
\end{equation}
for widely separated points, implying that the fluctuations do average out in the thermodynamic limit leaving us with a \emph{homogeneous} system of finite mean density. Hence the system is \emph{not} phase separated in the thermodynamic limit, even though on scales smaller than the crossover length ($L_N<\xi_{\ast}$) one does see dynamic hierarchical clusters violating Porod's law and a cusp in the equal time density correlator \cite{mishra2006active}, two hallmarks of fluctuation-dominated phase ordering. Eventually a large enough sytem will instead self-organize into a sort of critical phase with power law correlations in both the density and the order parameter. In contrast to generic scale invariance obtained for conserved dynamics in an anisotropic nonequilibrium steady state \cite{grinstein1990conservation}, no anisotropy survives at long distances here and the mechanism for self-organized criticality in the active nematic is different.

The presence of highly correlated fluctuations leads to non-standard scaling of the density distribution. The higher moments of the number fluctuations can be shown to scale as $\langle\delta N^k\rangle\propto N^{k(1-\eta(\Delta)/2)}$ (i.e. there is no multi-scaling). However, in the language of lattice-gas models, if we discretize and write $s_i=0,1$ as an occupation number within a small sub-volume indexed by $i$, then we have
\begin{equation}
	z=\dfrac{L^{\eta(\Delta)}}{\Delta^{1/2}}\left(\dfrac{1}{L^2}\sum_{i=1}^{L^2}s_i-\rho_0\right)\ ,
\end{equation}
with $\rho_0$ being the mean density, as the relevant scaling variable with a non-trivial limiting distribution (the probability distribution of $\rho=\sum_i s_i/L^2$ itself is sharply peaked around $\rho_0$ and not broad in the thermodynamic limit). As $L\to\infty$, Prob$(z)$ approaches a non-gaussian distribution whose cumulant generating function is
\begin{equation}
	\lim_{L\to\infty}\ln\langle e^{kz}\rangle=k^2\mu_2+k^3\mu_3\Delta^{1/2}+k^4\mu_4\Delta+o(\Delta)\ ,
\end{equation}
where $\mu_2,\mu_3\cdots$ are finite constants independent of $L$ and $\Delta$. So for low enough noise, the appropriately scaled density distribution is always unimodal in the thermodynamic limit, ruling out phase separation, even the unconventional one of Das and Barma \cite{barma2008singular}. The fact that the active current $\b{j}_{\mathrm{curl}}=\alpha_0\del\cdot\b{Q}$ is non-vanishing in the ordered phase and is not a pure gradient \footnote{It is important to note that having $\del\times\b{j}_{\mathrm{curl}}\neq 0$ is not a necessary condition for negating phase separation in general, it just happens to be so in this case.}, unlike the case of passive sliders on a fluctuating surface, is crucially responsible for this behaviour.

\section{Discussion}
Continuum models have long provided universal and generic descriptions of active systems and are in principal powerful enough to capture many of the dramatic consequences of activity, ranging from long-ranged $2d$ polar order in moving flocks \cite{toner1995long} to motility induced phase separation in scalar non-aligning active matter \cite{wittkowski2014scalar}. The use of renormalization group and field theoretic techniques allows us to systematically address the effect of fluctuations and noise in active systems, bringing the paradigm of universality to bear upon these non-equilibrium systems. Unlike dynamical critical phenomena in equilbrium, where mode coupling nonlinearities do not affect equal-time correlators in the steady state \cite{hohenberg1977theory}, the breaking of detailed balance in an active system encoded in the non-variational nature of the dynamics leads to a whole slew of rich phenomena, some of which we have tried to address in this paper.

In $2d$ at equilibrium, both polar and nematic liquid crystals or magnets have the same long-wavelength static description, that of the XY model. When active, the nematic system is distinctly different from its polar counterpart. Analyzing the symmetry in detail, we write down the leading order nonlinearities that are important and find them to all be marginal at the linear fixed point. The fields themselves being marginal, we find an infinite spectrum of marginal nonlinear terms, with most of them involving anisotropic couplings. The true symmetry of a nematic liquid crystal being a combined rotation of both the director and the spatial coordinates forces \emph{all} anisotropic couplings, linear and nonlinear, to be marginally irrelevant. Though all the anisotropic terms (including the active terms) flow to zero, we do not obtain an equilibrium nematic. Instead we find that the active current is dangerously irrelevant, by virtue of which the giant number fluctuations so engendered just get suppressed in a non-universal fashion, still violating the central limit theorem. This direct consequence of rotational symmetry of the model constrains the long-distance behaviour of the structure factor, forcing it to decay as a power law in distance, thereby ruling out the possibility of phase separation in the thermodynamic limit.

The absence of long-distance anisotropy also leads to the $2d$ active nematic only displaying quasi-long ranged order in the thermodynamic limit, making the bulk ordered state a critical phase with power law correlations in both the density and the nematic order parameter. Despite this disappointing result, we show that one can expect strong finite size effects when the active drive is stronger than the noise. In this case the nematic order parameter decays more slowly than a power law upto a crossover length scale, above which we recover QLRO once again. We also argue that the ordered nematic phase in both $2d$ active nematic and self-propelled rod systems must have the same universal description, and hence one cannot have long-ranged nematic order in \emph{any} locally driven $2d$ nematic (in the absence of long ranged interactions or hydrodynamics). Reconciling this result with previous numerical and experimental findings of long-ranged nematic order in self-propelled rod systems \cite{PhysRevE.95.020601,ginelli2010large} remains a theoretical challenge. By conventional expectations of universality and hydrodynamics, a simple resolution to this question, other than a long crossover, seems to be ruled out at least at the perturbative level.


\section{Acknowledgments}
We thank Mustansir Barma for useful discussions. This work was supported by the National Science Foundation at Syracuse University through awards DMR-1609208 (MCM,SS) and DGE-1068780 (MCM) and at KITP under the grant No. NSF PHY-1125915. SS and MCM thank the KITP for its hospitality during the completion of some of the work and the Syracuse Soft Matter Program for support. SR was supported by a J C Bose Fellowship of the SERB, Government of India and a Homi Bhabha Chair Professorship of the Tata Education and Development Trust.
\appendix
\section{Leading correction to anisotropic couplings}
\label{sec:aniso}

Considering just the interaction of the Nambu-Goldstone mode, we extend the simple analysis done in the main text and show how all the anisotropic terms have the same leading fluctuation correction. Taking as before $\alpha,\lambda\ll K$, we can neglect cross correlations in $\theta$ and $\delta\rho$, resulting in a factored gaussian distribution at the linear fixed point. We expand the trigonometric functions for small $\theta\ll1$,
\begin{align}
	\cos2\theta&=1-2\theta^2+o(\theta^3)\ ,\\
	\sin2\theta&=2\theta-\dfrac{4}{3}\theta^3+o(\theta^4)\ .
\end{align}
Systemizing the procedure, we first look at the diffusion anisotropy $\delta D\Gamma\cdot(\hat{\chi}\delta\rho)$ which was described in the main text as well,
\begin{equation}
	\delta D\Gamma\cdot(\hat{\chi}\delta\rho)=\delta D\left[\Gamma_1\delta\rho+\Gamma_2(2\delta\rho\theta)-\Gamma_1(2\delta\rho\theta^2)+\cdots\right]\ .
\end{equation}
As both the fields $\theta$ and $\delta\rho$ essentially behave as independent gaussian random variables at this order of approximation, we split the Nambu-Goldstone mode into slow and fast components $\theta=\theta_<+\theta_>$ and average over the short scale fluctuations,
\begin{equation}
	\langle\theta_>^2\rangle=\dfrac{\Delta}{4\pi K}\ln b\ .
\end{equation}
Here the average is performed in a thin momentum shell $\Lambda/b<|\b{q}_>|<\Lambda$ and $\delta D,\delta K$ only provide higher order corrections to the average. Using Wick's theorem and some simple combinatorics, we then get
\begin{align}
	\langle\Gamma_1(\delta\rho\theta^2)\rangle_>&=\langle\theta_>^2\rangle\Gamma_1\delta\rho\ ,\\
	\langle\Gamma_2(\delta\rho\theta^3)\rangle_>&=3\langle\theta_>^2\rangle\Gamma_2(\delta\rho\theta_<)\ .
\end{align}
One can similarly work out a similar calculation for the full trigonometric function, though we get the correct result from just looking at the first two terms as well.
\begin{equation}
	\delta D\langle\Gamma\cdot(\hat{\chi}\delta\rho)\rangle_>=\delta D(1-2\langle\theta_>^2\rangle)\Gamma\cdot(\hat{\chi}_<\delta\rho)\ ,
\end{equation}
where $\hat{\chi}_<=(\cos2\theta_<,\sin2\theta_<)$. So we immediately find, as mentioned in the main text, that the impact of the short scale director phase fluctuations is to renormalize the anisotropic coupling as
\begin{equation}
	\alpha'=\alpha(1-2\langle\theta_>^2\rangle)\implies\dfrac{\dd\alpha}{\dd\ell}=-\eta(\Delta)\alpha\ ,
\end{equation}
where $\eta(\Delta)=\Delta/2\pi K$. Similarly, doing the same for both $\delta K\hat{\chi}\cdot\Gamma\theta$ and $\lambda\hat{\chi}\times\Gamma\delta\rho$, we get the same result.
\begin{align}
	\delta K'&=\delta K(1-2\langle\theta_>^2\rangle)\ ,\\
	\lambda'&=\lambda(1-2\langle\theta_>^2\rangle)\ .
\end{align}
For the active current term $\alpha\Gamma\cdot\hat{\chi}$, expanding for small $\theta\ll1$, we have
\begin{equation}
	\alpha\Gamma\cdot\hat{\chi}=\alpha\left[\Gamma_22\theta-\Gamma_12\theta^2-\dfrac{4}{3}\Gamma_2\theta^3+\dfrac{4}{6}\Gamma_1\theta^4+\cdots\right]\ .
\end{equation}
Proceeding as before, we can replace $\langle\theta^3\rangle_>\to3\theta_<\langle\theta_>^2\rangle$ and $\langle\theta^4\rangle_>\to {}^4C_2\ \theta^2_<\langle\theta_>^2\rangle$ (where we have disregarded additive constants as all the angle terms come under derivatives). Working out the numbers, once again we get
\begin{equation}
	\alpha'=\alpha(1-2\langle\theta_>^2\rangle)\ .
\end{equation}
These were all the anisotropic terms that contribute at the level of linear hydrodynamics. We can follow the same procedure to show that the argument works even for higher order anisotropic nonlinearities, for example $K_n\hat{\vec{v}}\cdot\del\theta$. This term generates the KPZ like anisotropic nonlinearity $\sim\partial_x\theta\partial_y\theta$ at lowest order.
\begin{align}
	K_n&\left[(\del\cdot\hat{\chi})\partial_x\theta+(\del\times\hat{\chi})\partial_y\theta\right]=\nonumber\\
	&2K_n\left[2\partial_x\theta\partial_y\theta-\partial_x\theta^2\partial_x\theta+\partial_y\theta^2\partial_y\theta+\cdots\right]\ .
\end{align}
As before upon averaging we have, $\langle\theta^3\rangle_>\to3\theta_<\langle\theta_>^2\rangle$, and
\begin{equation}
	K_n\langle\hat{\vec{v}}\cdot\del\theta\rangle_>=K_n(1-2\langle\theta_>^2\rangle)(2\partial_x\theta_<\partial_y\theta_<+\cdots)\ ,
\end{equation}
implying as before that the coupling constant gets renormalized as $K_n'=K_n(1-2\langle\theta_>^2\rangle)$. The argument also applies to the advective coupling $D_n\hat{\vec{v}}\cdot\del\delta\rho$ term in the density equation, the calculation being entirely analogous. Note that unlike the regular KPZ nonlinearity $|\del\theta|^2$ which is marginally relevant in two spatial dimensions \cite{kardar1986dynamic}, the anisotropic version present here is always marginally irrelevant due to rotational symmetry. The usual KPZ nonlinearity is also forbidden in the density equation as it is not a total divergence and in the phase equation as it violates parity. A similar term $|\del\delta\rho|^2$ is also forbidden in both equations for the same reasons. There are many other anisotropic nonlinearities that also occur in an equilibrium lyotropic nematic, and hence such terms will automatically be generated after an iteration of the coarse-graining procedure. Once generated though, these terms will be subject to the same analysis done above in subsequent iterations of the renormalization group flow. So if we begin with all these higher order anisotropic nonlinearities being small, they remain so at least to leading order, flowing to zero for any non-zero noise.

\section{Renormalization group flow equations}
\label{sec:renorm}
\begin{figure}[h]
	\centering
	\includegraphics[width=0.5\textwidth]{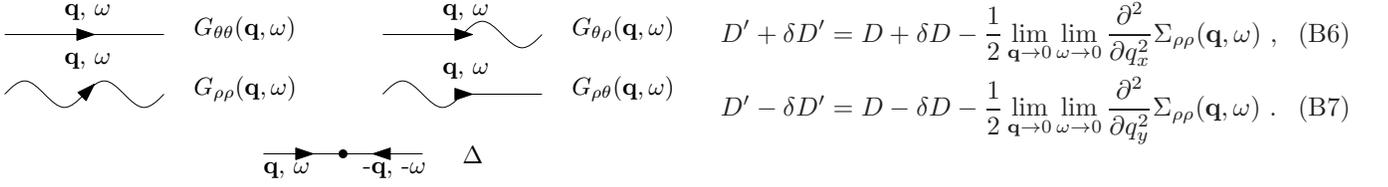}
	\caption{The field propagators and noise vertex}
	\label{fig:prop}
\end{figure}

\begin{figure*}[]
	\centering
	\includegraphics[width=\textwidth]{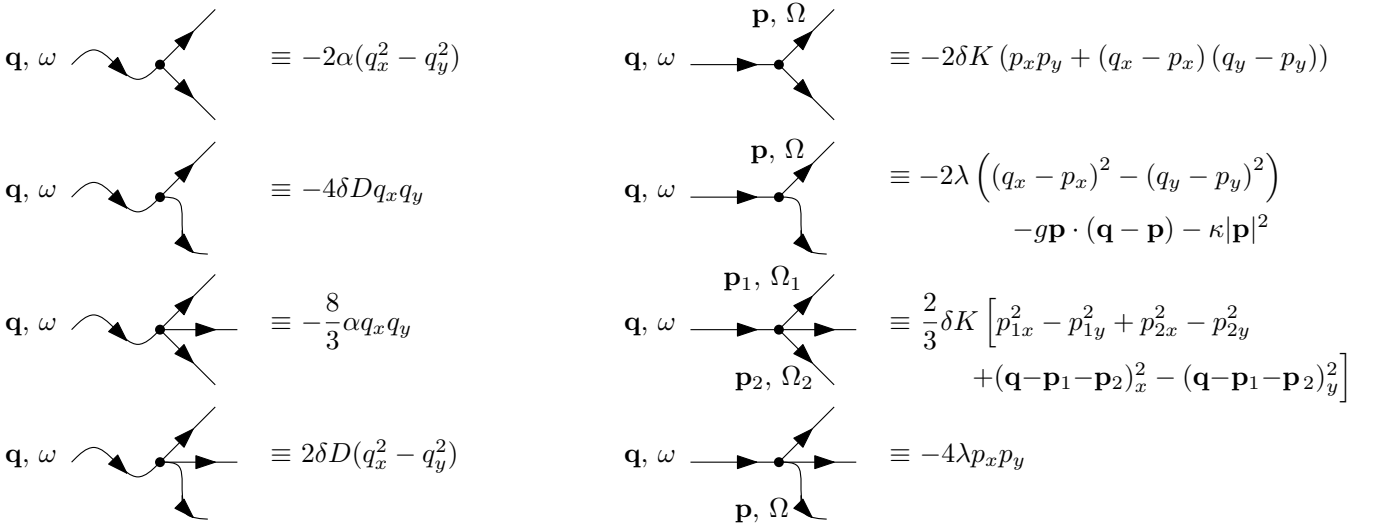}
	\caption{The interaction vertices}
	\label{fig:intvert}
\end{figure*}

Using a diagrammatic approach, the propagators and the noise vertex are drawn in Fig.~\ref{fig:prop} and the list of leading order interaction vertices as given in Eqs.~\ref{eq:rhoeqn} and \ref{eq:thetaeqn} are drawn in Fig.~\ref{fig:intvert} (the cubic vertex $\sim\sigma\del^2\delta\rho^2$ is not shown as it turns out to not contribute at lowest order). Upon including the interactions, the renormalized propagator $G_{\mathrm{R}}$ satisfies the Dyson equation
\begin{equation}
	G_{\mathrm{R}}^{-1}(\b{q},\omega)=G^{-1}(\b{q},\omega)-\Sigma(\b{q},\omega)\ ,
\end{equation}
where $\Sigma(\b{q},\omega)$ is the interaction ``self-energy'' given by the sum of all one-particle irreducible diagrams (1PI).

To first order in the noise variance ($\mathcal{O}(\Delta)$), only cubic and quartic vertices contribute to the self-energy. Having split the fields into slow and fast components ($\delta\rho=\delta\rho_{<}+\delta\rho_{>}$ and $\theta=\theta_{<}+\theta_{>}$) and averaging over $\delta\rho_{>}$ and $\theta_{>}$ using the noise, we can write down the corrected linear couplings as follows
\begin{align}
	K_3'&=K_3-\dfrac{1}{2}\lim_{\b{q}\to 0}\lim_{\omega\to 0}\dfrac{\partial^2}{\partial q_x^2}\Sigma_{\theta\theta}(\b{q},\omega)\ ,\\
	K_1'&=K_1-\dfrac{1}{2}\lim_{\b{q}\to 0}\lim_{\omega\to 0}\dfrac{\partial^2}{\partial q_y^2}\Sigma_{\theta\theta}(\b{q},\omega)\ ,\\
	\lambda'&=\lambda+\dfrac{1}{2}\lim_{\b{q}\to 0}\lim_{\omega\to 0}\dfrac{\partial^2}{\partial q_x\partial q_y}\Sigma_{\theta\rho}(\b{q},\omega)\ ,\\
	\alpha'&=\alpha+\dfrac{1}{4}\lim_{\b{q}\to 0}\lim_{\omega\to 0}\dfrac{\partial^2}{\partial q_x\partial q_y}\Sigma_{\rho\theta}(\b{q},\omega)\ ,\\
	D'+\delta D'&=D+\delta D-\dfrac{1}{2}\lim_{\b{q}\to 0}\lim_{\omega\to 0}\dfrac{\partial^2}{\partial q_x^2}\Sigma_{\rho\rho}(\b{q},\omega)\ ,\\
	D'-\delta D'&=D-\delta D-\dfrac{1}{2}\lim_{\b{q}\to 0}\lim_{\omega\to 0}\dfrac{\partial^2}{\partial q_y^2}\Sigma_{\rho\rho}(\b{q},\omega)\ .
\end{align}
$K_3=K+\delta K$ and $K_1=K-\delta K$ are the bend and splay elastic constants respectively. The correction to the noise vertex is given by the sum over bubble diagrams $\Pi_{\theta\theta}(\b{q},\omega)$,
\begin{equation}
	\Delta'=\Delta+\lim_{\b{q}\to0}\lim_{\omega\to0}\Pi_{\theta\theta}(\b{q},\omega)\ .
\end{equation}
With this we can proceed to compute the full renormalization group flow equations. After a total of about $20$ loop integrals for the self energy and $4$ for the noise vertex corrections, both $\Sigma(\b{q},\omega)$ and $\Pi_{\theta\theta}(\b{q},\omega)$ to lowest order in wavevector $\b{q}$ and at zero frequency ($\omega=0$) are found to be
\begin{widetext}
	\vspace{-1.5em}
\begin{align}
	\Sigma_{\rho\rho}(\b{q},\omega=0)&=\dfrac{\Delta}{2\pi K}\ln b\left\{-\dfrac{\alpha\kappa}{K}+\delta D\left[1+\dfrac{\alpha\lambda}{K(1+K)}+\left(\dfrac{\delta K}{K}\right)^2\right]\right\}(q_x^2-q_y^2)\ ,\\
	\Sigma_{\theta\theta}(\b{q},\omega=0)&=\dfrac{\Delta}{2\pi K}\ln b\left\{-\left[\dfrac{\delta K^2(K^3+5K^2-5K-1)+4\alpha\lambda K(1+3K+4K^2)}{2K(1+K)^3}\right]q^2\right.\nonumber\\
																																   &\quad\qquad\quad+\left.\left[\delta K+g\alpha\dfrac{1+3K+4K^2}{K(1+K)^3}-\dfrac{2K^2\alpha\kappa}{(1+K)^3}\right](q_x^2-q_y^2)\right\}\ ,\\
	\Sigma_{\theta\rho}(\b{q},\omega=0)&=\dfrac{\Delta}{2\pi}\ln b\left[-2\dfrac{\lambda}{K}-\dfrac{\delta K\kappa}{2K^2}-\dfrac{2K\delta D\kappa}{(1+K)^3}+\dfrac{g}{2K^2}\left(\dfrac{2K\delta D(1+3K+4K^2)}{(1+K)^3}-\delta K\right)\right.\nonumber\\
																																	   &+\left.\dfrac{\alpha\kappa}{4K^2(1+K)^3}\left((1+4K-K^2)(\kappa-g)+\dfrac{g(K-1)(1+3K)^2+\kappa(1+5K+11K^2-K^3)}{1+K}\right)\right]q_xq_y\ ,\label{eq:sigmalambda}\\
	\Sigma_{\rho\theta}(\b{q},\omega=0)&=-\dfrac{\alpha\Delta}{2\pi K}\ln b\left[1+\dfrac{2(1+K)^2\delta K^2+4K(1+K)\alpha\lambda+K(1+4K)\delta K\delta D+K^2\delta D^2}{2K^2(1+K)^2}\right]4q_xq_y\ ,\\
	\Pi_{\theta\theta}(\b{q}=\b{0},\omega=0)&=\dfrac{\Delta^2}{2\pi}\ln b\left[2\dfrac{\delta K^2}{K^3}+\delta K\alpha(\kappa-g)\dfrac{1+2K}{K^3(1+K)^2}+\dfrac{\alpha^2(g^2+\lambda^2)}{K^3(1+K)^3}(1+3K+K^2)\right]\ .
\end{align}
\end{widetext}
As mentioned in the main text, the leading correction to the diffusion constant $D$ comes from the $\sigma\del^2\delta\rho^2$ term, contributing only at order $\mathcal{O}(\Delta\sigma\alpha\delta D)$, which is subdominant to lower order corrections in other terms and is itself irrelevant as it involves both $\alpha$ and $\delta D$. So we keep $D=1$ fixed by setting $z=2$. $\sigma$ also renormalizes only rather weakly with the leading vertex correction being $\mathcal{O}(\Delta\sigma^2\alpha^2)$ and hence we don't worry about it any further by setting $\sigma=0$. The only other computation left is that of the vertex correction to $g$ and $\kappa$. To lowest order this involves five diagrams. The vertex itself is constrained by rotational symmetry and is generally given as
\begin{figure}[]
	\centering
	\includegraphics[width=0.3\textwidth]{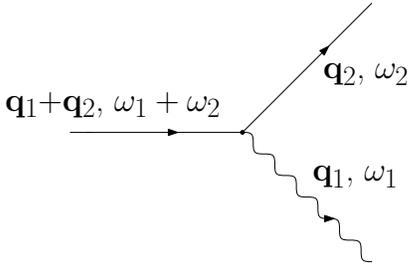}
	\caption{The cubic vertex $V_{\theta\rho\theta}$ involving the isotropic nonlinearities $g$ and $\kappa$. The incoming straight line is a director fluctuation which decays into a density mode (the wavy line) with wavevector $\b{q}_1$ and another director fluctuation mode with wavevector $\b{q}_{2}$.}
	\label{fig:vertex}
\end{figure}
\begin{gather}
		V_{\theta\rho\theta}(\b{q}_1,\b{q}_2)=-2\lambda(q_{1x}^2-q_{1y}^2)-g\b{q}_1\cdot\b{q}_2-\kappa|\b{q}_2|^2\nonumber\\
		\quad\qquad-\gamma_1(q_{2x}^2-q_{2y}^2)-\gamma_2(q_{1x}q_{2x}-q_{1y}q_{2y})-\gamma_3q_{2x}q_{2y}\ ,
\end{gather}
for vanishing outgoing frequencies ($\omega_{1,2}=0$) and with $\b{q}_1$ and $\b{q}_2$ as the outgoing wavevectors in the density and phase modes (see Fig.~\ref{fig:vertex}). The $\gamma_{1,2,3}$ couplings are permitted by symmetry and will in general be generated upon renormalization. These terms are all anisotropic in nature and arise from the density dependence of the Frank constant anisotropy ($\gamma_1\delta\rho\Gamma_1\theta$) or from the KPZ like advective nonlinearity $K_n\vec{v}\cdot\del\theta$ which when expanded leads to both $\gamma_2(\partial_x\delta\rho\partial_x\theta-\partial_y\delta\rho\partial_y\theta)$ and $\gamma_3\partial_x\theta\partial_y\theta$ terms. Using the argument in Appendix~\ref{sec:aniso} used for all the anisotropic terms, we conclude $\gamma_{1,2,3}$ to all be irrelevant and do not consider them any further. The leading correction of the cubic vertex gives the recursion relations for $g$, $\kappa$ and $\lambda$. Importantly, as required by rotational invariance, no corrections $\propto |\b{q}_1|^2$ or $q_{1x}q_{1y}$ arise in $V_{\theta\rho\theta}$, with the corresponding loop integral contributions cancelling only after summing over all the leading diagrams. Having already computed the loop correction to $\lambda$ from the renormalization of the $G_{\theta\rho}$ propagator, the two flow equations \emph{must} coincide as the the coefficient in the cubic vertex is related to the linear coupling by rotational symmetry. This requirement will allow us to fix the value of the yet unknown scale factor $\zeta_{\theta}$ for the slow angle field $\theta_{<}$. The corrections to $g$, $\kappa$ and $\lambda$ from the vertex are

\begin{gather}
	g'=g-4\delta D\lambda\Delta\dfrac{1}{2\pi}\ln b\dfrac{1+3K+4K^2}{K(1+K)^3}\ ,\\
	\kappa'=\kappa-2\delta D\lambda\Delta\dfrac{1}{2\pi}\ln b\dfrac{1+3K+4K^2}{K(1+K)^3}\ ,\\
	\lambda'=\lambda-\dfrac{\Delta}{2\pi K}\ln b\left[\lambda-g\delta D\dfrac{1+3K+4K^2}{2(1+K)^3}+\dfrac{g\delta K}{4K}\right.\nonumber\\
	\quad\qquad\quad\left.+\dfrac{\delta D\kappa K^2}{(1+K)^3}+\dfrac{\delta K\kappa}{4K}\right]\ .\label{eq:vertlambda}
\end{gather}
Having integrated out a thin shell of short scale fluctuations ($\delta\rho_{>},\theta_{>}$), we now rescale back both space and time ($\vec{r}\to\vec{r}b$, $t\to tb^z$) along with the slow fields ($\delta\rho_<\to\delta\rho_<b^{\zeta_{\rho}}$ and $\theta_<\to\theta_<b^{\zeta_{\theta}}$) in order to restore the cutoff back to $\Lambda$. Having already set $z=2$, writing $\ell=\ln b\ll1$, we obtain differential recursion relations for the various coupling constants.

\begin{widetext}
	\vspace{-1.5em}
	\begin{align}
	\dfrac{\dd\delta D}{\dd\ell}&=\dfrac{1+K}{2}\bar{\kappa}-\delta D\dfrac{\bar{\lambda}}{2}-\dfrac{\Delta\delta D}{2\pi K}\left[1+\left(\dfrac{\delta K}{K}\right)^2\right]\ ,\\
	\dfrac{\dd K}{\dd\ell}&=K\left\{c_K\bar{\lambda}+d_K\dfrac{\Delta}{4\pi K}\left(\dfrac{\delta K}{K}\right)^2\right\}\ ,\\
	\dfrac{\dd \delta K}{\dd\ell}&=-\dfrac{\Delta}{2\pi K}\delta K+K\left(-\bar{g}\dfrac{c_K}{2}+\bar{\kappa}\dfrac{K^2}{(1+K)^2}\right)\ ,\\
	\dfrac{\dd\Delta}{\dd\ell}&=\Delta\left\{-2\zeta_{\theta}+(\bar{\kappa}-\bar{g})\delta K\dfrac{(1+2K)}{2K(1+K)}+\dfrac{\Delta}{\pi K}\left(\dfrac{\delta K}{K}\right)^2\right\}\ ,\\
	\dfrac{\dd\lambda}{\dd\ell}&=\lambda\left(\zeta_{\rho}-\zeta_{\theta}-\dfrac{\Delta}{2\pi K}\right)+\dfrac{g\Delta}{8\pi K^2}\left(2\delta D\dfrac{Kc_K}{1+K}-\delta K\right)-\dfrac{\kappa\Delta}{8\pi K^2}\left(4\delta D\dfrac{K^3}{(1+K)^3}+\delta K\right)\ ,\label{eq:lambda}\\
	\dfrac{\dd \alpha}{\dd\ell}&=\alpha\left\{\zeta_{\theta}-\zeta_{\rho}-\bar{\lambda}-\dfrac{\Delta}{2\pi K}\left[1+\dfrac{\delta D^2}{2(1+K)^2}+\delta D\delta K\dfrac{(1+4K)}{2K(1+K)^2}+\left(\dfrac{\delta K}{K}\right)^2\right]\right\}\ ,\\
	\dfrac{\dd g}{\dd\ell}&=g\zeta_{\rho}-2\Delta\delta D\lambda\dfrac{c_K}{\pi K(1+K)}\ ,\\
	\dfrac{\dd\kappa}{\dd\ell}&=\kappa\zeta_{\rho}-\Delta\delta D\lambda\dfrac{c_K}{\pi K(1+K)}\ .
	\end{align}
\end{widetext}

In order to simplify the notation we have used
\begin{equation}
	d_K=\dfrac{(K-1)(1+6K+K^2)}{(1+K)^3}\ ,
\end{equation}
along with $c_K=(1+3K+4K^2)/(1+K)^2$ as given in the main text, and $\bar{g}=\alpha g\Delta/[\pi K^2(1+K)]$ and $\bar{\kappa}=\alpha\kappa\Delta/[\pi K^2(1+K)]$ are defined similar to $\bar{\lambda}$. As mentioned before, comparing the recursion relations for $\lambda$ independently obtained from both Eq.~\ref{eq:sigmalambda} and Eq.~\ref{eq:vertlambda} we obtain $\zeta_{\theta}=0$ to lowest order.

One can easily check that $\delta K=\delta D=g=\kappa=0$ provides an invariant subspace in $\{K,\bar{\lambda}\}$ with the noise variance $\Delta$ unrenormalized. Looking at small deviations from this subspace, we linearize the recursion relations around a particular $\{K(\ell),\bar{\lambda}(\ell)\}$ trajectory. To linear order in $\delta K,\delta D,g$ and $\kappa$, the flow of $K,\Delta$ and $\bar{\lambda}$ remains unchanged, so Eqs.~\ref{eq:Kflow} and~\ref{eq:lambdabarflow} continue to hold even for small transverse deviations from the invariant submanifold. The linearized flow equations are
\begin{align}
	\dfrac{\dd\overline{\delta D}}{\dd\ell}&=\dfrac{\bar{\kappa}}{2}-\overline{\delta D}\left[\eta(\Delta)+\bar{\lambda}\left(\dfrac{Kc_K}{1+K}+\dfrac{1}{2}\right)\right]\ ,\\
	\dfrac{\dd\overline{\delta K}}{\dd\ell}&=-\dfrac{c_K}{2}\bar{g}+\dfrac{\bar{\kappa}K^2}{(1+K)^2}-\overline{\delta K}\left(\eta(\Delta)+c_K\bar{\lambda}\right)\ ,\\
	\dfrac{\dd\bar{g}}{\dd\ell}&=-\bar{\lambda}\dfrac{2\Delta c_K}{\pi K}\overline{\delta D}-\bar{g}\left(\eta(\Delta)+b_K\bar{\lambda}\right)\ ,\\
	\dfrac{\dd\bar{\kappa}}{\dd\ell}&=-\bar{\lambda}\dfrac{\Delta c_K}{\pi K}\overline{\delta D}-\bar{\kappa}\left(\eta(\Delta)+b_K\bar{\lambda}\right)\ .
\end{align}
Once again $b_K=1+c_K(2+3K)/(1+K)$ and $\eta(\Delta)=\Delta/2\pi K$ as given in the main text. We also normalize the diffusion and elastic anisotropies as $\overline{\delta D}=\delta D/(1+K)$ and $\overline{\delta K}=\delta K/K$. Writing the above equations as $\dd\Phi/\dd\ell=\mathsf{L}\Phi$ where $\Phi=\{\overline{\delta D},\overline{\delta K},\bar{g},\bar{\kappa}\}$, we treat $K,\Delta$ and $\bar{\lambda}$ as essentially constant at a given point on the renormalization flow trajectory. Diagonalizing the linear matrix $\mathsf{L}$, the corresponding eigenvalues $y_i$ ($i=1$ to $4$) control the scaling dimensions and (ir)relevance of the various couplings. The first two eigenvalues are
\begin{align}
	y_1&=-(\eta(\Delta)+b_K\bar{\lambda})\ ,\\
	y_2&=-(\eta(\Delta)+c_K\bar{\lambda})\ ,
\end{align}
both of which are always negative for $\bar{\lambda}>0$. We will primarily focus only on the first quadrant of the $\{K,\bar{\lambda}\}$ plane, though the basin of stability of the line of fixed points on the $K$-axis, extends to a small region of $\bar{\lambda}<0$, which becomes vanishingly small for large $K$ and small $\Delta$. Outside this region (i.e. for $\bar{\lambda}<0$), the flow is perturbatively unstable even within the $\{K,\bar{\lambda}\}$ plane, and we don't address it any further. As both $y_{1,2}<0$, both these directions are stable and flow to zero at a fixed point with finite $K$ and $\Delta$. The other two eigenvalues of $\mathsf{L}$ are a complex conjugate pair,
\begin{equation}
	y_{3,4}=-\eta(\Delta)-A_0\dfrac{\bar{\lambda}}{2}\pm\dfrac{1}{2}\sqrt{\bar{\lambda}^2A_1^2-\bar{\lambda}\dfrac{2c_K\Delta}{\pi K}}\ .
\end{equation}
$A_{0,1}=b_K\pm[Kc_K/(1+K)+1/2]$ are complicated functions of $K$, that remain bounded for all $0\leq K\leq\infty$. Importantly $A_0>0$ for all $K\geq 0$ and hence for small $\bar{\lambda}$, $\mathrm{Re}(y_{3,4})<0$ and $\mathrm{Im}(y_{3,4})=\pm\sqrt{\eta(\Delta)c_K\bar{\lambda}}+\mathcal{O}(\bar{\lambda})$. Even for larger $\bar{\lambda}$, one can show that $\mathrm{Re}(y_3)<0$ for all $\bar{\lambda}>0$ as long as $K>-1/3$. Hence in these directions as well, the flow is stable, though oscillatory. With this we conclude that the $\{K,\bar{\lambda}\}$ subspace is linearly stable and attracting in the top right quadrant, with deviations from it being irrelevant. Of course, all of this is only a perturbatively valid statement and for larger $\bar{\lambda}>0$, one could have a phase transition to a strong ``activity'' fixed point. Such a scenario though is currently inaccessible within a perturbative treatment.

Though our analysis was performed strictly in two dimensions, for $d=2+\epsilon$, we can also formally extend the recursion relations for small $\epsilon$ by just accounting for the dimensional change in couplings while keeping the loop corrections the same as for $d=2$. Doing so for the simple case when $\delta K=\delta D=g=\kappa=0$, we find
\begin{align}
	\dfrac{\dd T}{\dd\ell}&=-T(\epsilon+c_K\bar{\lambda})\ ,\\
	\dfrac{\dd\bar{\lambda}}{\dd\ell}&=-\bar{\lambda}(\epsilon+T+b_K\bar{\lambda})\ ,
\end{align}
where $T=\Delta/\pi K$ is just a scaled noise variance. For $\epsilon>0$, we immediately find that both the effective noise and the activity are driven to zero very quickly, making all the nonlinearities irrelevant. For $\epsilon<0$, there is a fixed point in $(T,\bar{\lambda})$ of $\mathcal{O}(\epsilon)$, but the only physical dimension below two is $d=1$, in which a nematic doesn't break a continuous symmetry. So above $d=2$, activity is always irrelevant (dangerously though as it still causes large number fluctuations) and we recover the linearized description of an active nematic.


\end{document}